\documentclass[12pt]{article}%
\usepackage{cite}
\usepackage{fix-cm}
\usepackage{heck}
\usepackage{graphicx}
\usepackage{multicol}
\usepackage{amsmath}
\usepackage{amsfonts}
\usepackage{amssymb}
\usepackage{hyperref}
\usepackage{setspace}
\usepackage[all]{xy}
\usepackage{verbatim}
\usepackage{color}
\usepackage{epsfig}
\usepackage{mathtools}
\providecommand{\U}[1]{\protect\rule{.1in}{.1in}}
\numberwithin{equation}{section}

\newcommand{\Tr}{\, {\rm Tr}}

\def\be{\begin{equation}}
\def\ee{\end{equation}}
\def\bea{\begin{eqnarray}}
\def\eea{\end{eqnarray}}
\def\spc{\hspace{1pt}}

\def\ppi{\pi}
\def\sN{{{}_{\! N\smpc}}}

\def\ellcc{\ell}
\def\delbar{\overline{\partial}}

\newcommand{\smpc}{\hspace{.5pt}}

\def\xx{{\mbox{\small $\hat{X}$}}} 

\def\xxn{y}
\def\is{\! & \! \! = \! & \! \! }

\def\ccdot{\! \cdot \! \spc}
\def\aA{{\mbox{\fontsize{6.5}{7}\selectfont$A$}}}

\def\vvv{{\mbox{\fontsize{10}{11}\selectfont$v$}}}

\def\Dbar{\overline{D}}

\def\bA{\mathcal{A}}
\def\tQ{\widetilde{Q}}
\def\qQ{Q}
\def\ra{\rangle}
\def\la{\langle}

\renewcommand{\footnotesize}{\small}
\begin{document}

\addtolength{\baselineskip}{.6mm}
\date{December 2011}

\title{Gravity Amplitudes\\[4mm] from a Gaussian Matrix Model}

\institution{IAS}{\centerline{${}^{1}$School of Natural Sciences, Institute for
Advanced Study, Princeton, NJ
08540, USA}}

\institution{PU}{\centerline{${}^{2}$Department of Physics, Princeton University,
Princeton, NJ 08544,
USA}}

\authors{Jonathan J. Heckman\worksat{\IAS}\footnote{e-mail: {\tt
jheckman@ias.edu}} and Herman
Verlinde \worksat{\PU}\footnote{e-mail: {\tt verlinde@princeton.edu}}}

\abstract{We reformulate MHV scattering amplitudes in 4D gauge theory and supergravity as
correlation functions of bilinear operators in a supersymmetric gaussian matrix model.
The model retains the symmetries of an $S^{4}$ of radius $\ellcc$ and
the matrix variables are represented as linear operators acting on a finite-dimensional
Hilbert space. Bilinear fields of the model generate a current algebra. In the large $N$ double scaling limit
where $\ell_{pl} \sim \ellcc / \sqrt{N}$ is held fixed, there is an emergent
flat 4D space-time with a built in short distance cutoff.}

\maketitle

\enlargethispage{\baselineskip}

\setcounter{tocdepth}{2}
\tableofcontents

\section{Introduction}

Recent progress in the study of scattering amplitudes in supersymmetric gauge theory and gravity has
revealed surprising structures that hint at the existence of deep new principles
\cite{ArkaniHamed:2009dn, Mason:2009qx, Alday:2010ku, CaronHuot:2010ek, Drummond:2010zv, Goncharov:2010jf, Alday:2010vh, ArkaniHamed:2010kv, Adamo:2011dq, Eden:2011yp, Eden:2011ku}.
This development was triggered by, and has gone hand in hand with, the exploration
of the various dualities between gauge theory, gravity and string theory -- most notably the AdS/CFT correspondence, and the proposed reformulation of ${\cal N}\! =\! 4$ SYM amplitudes via twistor string theory \cite{Witten:2003nn}, \cite{Berkovits:2004hg, Berkovits:2004jj}  (see \cite{Cachazo:2005ga} for an early review).

While influential in the beginning, the twistor string program has been relatively dormant in recent
years, in large part because of the realization that the theory contains conformal supergravity,
and by the recognized inconsistency of the latter. In spite of this apparent roadblock,
the twistor string is clearly a natural and beautiful idea that deserves to
find its place  among the realm of consistent string theories. Rather than a fatal weakness,
the unexpected emergence of gravity could count as an extra motive for trying to
get to the bottom of its true significance.

In the accompanying paper \cite{TMMII}, we propose a new interpretation of twistor string theory that
potentially avoids the trap that led to its apparent inconsistency. The new idea is to view the
strings as emergent low energy degrees of freedom of a pure holomorphic $U(N N_c)$ Chern-Simons gauge
theory in the presence of a $U(N)$ background flux with a large instanton number $k_N$ \cite{TMM}.\footnote{
Our choice to consider the large instanton limit was stimulated by previous proposals that
formulate theories of quantum gravity in terms of large $N$ matrices as for example in
\cite{Banks:1996vh, Ishibashi:1996xs, juanAdS, gkPol, witHolOne}.}
The twistor strings then arise as collective  modes of the instantons. Their dynamics is
captured by an effective theory that contains a $U(N_c)$ gauge field $\bA$, also with a hCS action,
living on a non-commutative twistor  space. In addition, the instantons give rise to a
pair of defect modes $\qQ$ and $\tQ$.  As shown in \cite{TMMII}, the low energy gauge
field and defect modes naturally combine into a matrix model that encompasses the ADHM
construction of instantons. The matrix model is initially formulated at finite $N$ and
comes with a length scale $\ellcc$ given by the radius of an $S^{4}$. See \cite{Lechtenfeld:2005xi}
for earlier work on potential connections between matrix models and twistor string theory.

As a concrete output of this new proposal, we have identified a simple gaussian large $N$ matrix
model given by an action of the form:
\begin{equation}
\label{gone}
S_{\text{MM}}(Q,\tQ) = \Tr \left(\tQ \overline{D}_{\mathcal{A}} Q \right)
\end{equation}
where $Q$ and $\tQ$ are finite matrices, and $\overline{D}_{\mathcal{A}}$ is a non-commutative covariant derivative,
that is, a linear operator acting on $Q$. This gaussian model is obtained by starting from  the interacting ADHM matrix model introduced in \cite{TMMII},
and working with respect to a fixed background for the non-commutative gauge field $\bA$. Our goal in this paper is to establish
that this truncated model describes the MHV sector of an emerging space-time theory.

The gaussian matrix model (\ref{gone}) enjoys a number of remarkable properties.
The model comes with an $SO(5)$ symmetry, that reflects its relation with an underlying
$S^{4}$ space-time geometry.  Even at finite $N$, the matrices acquire a natural geometric
interpretation as holomorphic functions  on twistor space. As a result, the basic link
 (see e.g. \cite{Penrose:1967wn, Penrose:1968me}) between twistors and space-time physics is preserved.
The flat space continuum limit is obtained by taking a double scaling limit $N \rightarrow \infty$ with:
\begin{equation}\label{doubscale}
\ell_{pl}^{2} = \frac{\ellcc^{2}}{N}
\end{equation}
held fixed. In this limit, we will identify spin one excitations for a $u(N_c)$ gauge theory as well as
spin two excitations corresponding to deformations of the space-time geometry. This motivates an
identification of the short distance cutoff $\ell_{pl}$ with the Planck length.

The basic link we establish in this paper is that in this double scaling limit, states with
specified momentum and helicity are represented by currents $\mathcal{J}$ made
from bilinears in the matrix variables. Correlation functions of multiple $\mathcal{J}_{i}$
insertions compute amplitudes of the 4D theory via the correspondence:
\begin{equation}
\text{Amplitude} = \Bigl \langle \mathcal{J}_{1} ... \mathcal{J}_{m} \Bigr \rangle_{\text{MM}}.
\end{equation}
This link between currents and space-time fields is somewhat similar to the AdS/CFT dictionary,
and to the map between target space fields and vertex operator of the string worldsheet theory.
MHV gauge theory amplitudes are computed via a $u(N_c)$ current algebra, and naturally
reproduce the Parke-Taylor  formula \cite{Parke:1986gb}.
Moreover, we find that the matrix model does not give rise to conformal gravity amplitudes,
but rather, in the strict large $N$, $\ell_{pl} \to 0$ limit, produces an effective action that
matches the generating functional of MHV amplitudes in ${\cal N}=4$ SYM theory
introduced by \cite{Boels:2006ir, Boels:2007qn}.

A striking feature of the matrix model is that it also allow for other geometric currents,
that, as long as $\ell_{pl}$ is kept finite, generate complex structure deformations of the non-commutative
twistor geometry. The properties of these currents and deformations are strongly reminiscent of the
non-linear graviton construction of Penrose \cite{Penrose:1976jq}. Motivated by this relationship,
we will compute correlation functions of these geometric currents and quite remarkably, we will find that
they reproduce the BGK amplitude for MHV graviton scattering \cite{Berends:1988zp}. In this sense,
we can identify a natural algebra associated with MHV gravity amplitudes, which is
also reminiscent of BCJ \cite{Bern:2008qj} (see also \cite{Monteiro:2011pc}).
The final representation (and several geometric and combinatorial steps in the derivation) of the amplitude takes the same form
as the formula derived in \cite{Mason:2008jy}
\begin{equation}
\label{BGK}
i \mathcal{M}_{\rm BGK}= \kappa^{n-2} \delta^{4}\Bigl(\sum_{i=1}^n p_{i}\Bigr)\; \frac{\left\langle n\spc 1\right\rangle ^{8}}{\la 1\spc n\! -\! 1\ra \la
n\! -\! 1\spc n\ra\la  n \spc 1\ra}\frac{1}{C(n)}\underset
{k = 2}{\overset{n-2}{%
{\displaystyle\prod}
}}\frac{[ \spc k \spc  |  p_{1}+...+p_{k - 1}\vert n\ra
}{\la k \spc n\ra }+P_{(2,...,n-1)}.
\end{equation}
where the sum $P_{2,...,n-1}$ is over all permutations of the plus helicity gravitons. Here,
$\kappa = \sqrt{16 \pi G_{N}}$, and $C(n) = \la 1 2 \ra \la2 3\ra \cdots \la n\!-\!1\spc  n \ra\la n 1\ra$
is the usual Parke-Taylor denominator.

Our plan in this paper will be to start from the gaussian matrix model, and to show how features of the 4D space-time theory are built up.
This can be viewed as a ``bottom up'' perspective on the matrix model and its connection to a potential theory of emergent space-time and gravity.
In the companion paper \cite{TMMII}, we provide a more UV motivated perspective on the proposal.

The rest of this paper is organized as follows. In section \ref{sec:GMM} we introduce the basic features of the gaussian matrix model which follow from \cite{TMM, TMMII}. Section \ref{sec:SYMMS} studies the symmetries and currents of the matrix model. In section \ref{sec:spacetime} we provide a space-time interpretation for the matrix model. This will motivate the conjecture that correlators of the matrix model are connected with 4D physics. In section \ref{sec:WARMUP} we study correlators for a fuzzy $\mathbb{CP}^{1}$. This is of interest in its own right, and will serve as a springboard for the full computation of the matrix model correlation functions. We formulate the calculation of scattering amplitudes in section \ref{sec:SCATT}, and show that MHV gluon correlators are naturally reproduced from a $u(N_c)$ current algebra. In section \ref{sec:MHVgrav} we compute MHV graviton scattering amplitudes.
Section \ref{sec:CONC} contains our conclusions. Some additional technical details are included in the Appendices.

\section{Gaussian Matrix Model} \label{sec:GMM}

In this section we introduce the supersymmetric gaussian matrix model. The matrix variables are represented as
linear operators acting on a finite dimensional Hilbert space, obtained by quantizing ${\cal N}=4$ supersymmetric twistor space $\mathbb{CP}^{3|4}$.

\subsection{Matrix Action}

The matrix model that we will study in this paper is given by the gaussian integral over a conjugate pair
of ${\cal N}\! =\!4$ supersymmetric matrices $\qQ$ and $\widetilde{\qQ}$, acting on a certain finite dimensional
vector space.  To assemble the ${\cal N}\!=\!4$ supermultiplets,
we introduce four anticommuting coordinates $\psi^{i}$, $i=1,..,4$ and their hermitian conjugates $\psi_i^\dag$,
 and write the two matrix variables as  superfields $\tilde{\qQ}(\psi,\psi^\dag)$ and $\qQ(\psi,\psi^\dag)$.
 We will assume that $\psi$ and $\psi^\dag$ satisfy the  algebra $\bigl\{  \psi^{i},\psi_{j}^{\dag}\bigr\}  = \delta_{j}^{i}
\text{.}$ Besides the anti-commuting coordinates $\psi^i$, we now also introduce four mutually commuting
matrices $Z^\alpha$, $\alpha=1,..,4$, of the same size as $\tilde{\qQ}$ and $\qQ$, and their hermitian conjugates $Z^\dag_\alpha$. We will specify the
precise form of these matrices $Z^\alpha$ momentarily. We combine the matrices $Z^\alpha$ and  anti-commuting variables $
\psi^i$ as
\bea
\mathcal{Z}^{I}\is(Z^{\alpha}|\psi^{i})
\eea
which can be viewed as a system of coordinates on $\mathbb{C}^{4|4}$.
The action for the gaussian matrix model now takes the following simple form \cite{TMMII}
\bea
\label{mmaction}
S_{\rm MM} (\tQ,\qQ) = \text{Tr} 
\Bigl( {\cal I}_{I\!\spc J} \tQ\mathcal{Z}^{I}\qQ\mathcal{Z}^{J}\Bigr)  .
\eea
Here ${\cal I}_{I\!\spc J}$ is a pairing which is anti-symmetric (resp. symmetric) on the bosonic (resp. fermionic) part of
$\mathbb{C}^{4|4}$, and ${\rm Tr}$ denotes the trace over
the supersymmetric vector space 
on which the matrix superfields act. We will specify this vector space below.

Now let us specify the four matrices $Z^\alpha$. Initially, we introduce the $Z^\alpha$ and their hermitian
conjugates $Z^\dag_\alpha$
as bosonic oscillators, which satisfy the canonical commutation relations $\bigl[  Z^{\alpha},Z_{\beta}^{\dag}\bigr]  =
\delta_{\beta}^{\alpha}$.
The representation space of this algebra looks like the Hilbert space of four simple harmonic oscillators, on which
the $Z^\alpha$ act like annihilation operators and $Z^\dag_\beta$ act as creation operators. Summarizing, the oscillator algebra is:
\bea
\label{zcom}
\bigl[  Z^{\alpha},Z_{\beta}^{\dag}\bigr]  = \delta_{\beta}^{\alpha}\,  \quad &; &  \quad \bigl\{  \psi^{i},\psi_{j}^{\dag}\bigr
\}  = \delta_{j}^{i}
\eea
which gives an oscillator algebra representation of $gl(4|4,\mathbb{C})$. This contains $psl(4|4,\mathbb{C})$, the complexified superconformal algebra.
We denote by $\mathcal{H}_{\mathbb{C}^{4|4}}$ the Fock space of states generated by the creation
operators  $\mathcal{Z}_{J}^{\dag}$. The Hilbert space $\mathcal{H}_{\mathbb{C}^{4|4}}$
is the linear space spanned by all the fuzzy points on a non-commutative
$\mathbb{C}^{4|4}$ \cite{FUZZ,TMM}. Each basis state in the Fock space represents one Planck
cell of the non-commutative space, and since $\mathbb{C}^{4|4}$ is non-compact, the
associated Hilbert space  $\mathcal{H}_{\mathbb{C}^{4|4}}$ is infinite dimensional.

To make the Hilbert space finite dimensional, we will now take the K\"ahler quotient
\bea
\label{kquo}
\mathbb{C}^{4|4}/\!/ U(1) \is \mathbb{CP}^{3|4},
\eea
where the $U(1)$ acts by uniform phase rotation on all supercoordinates ${\mathcal Z}^I$.
The projective space $\mathbb{CP}^{3|4}$ is compact, and its non-commutative realization has a finite number of
Planck cells.
We should thus expect to find a finite  dimensional Hilbert space. The $U(1)$ symmetry, that features in the K\"ahler
quotient (\ref{kquo}), is generated by the homogeneity operator
\bea
\frak{D}_0 \is {Z}_{\alpha}^\dag{Z}^{\alpha} + \psi^\dag_i \psi^i.
\eea
This operator has an integer spectrum, given by the sum of the ${\cal Z^I}$ oscillator levels.
To perform the K\"ahler quotient we consider eigenstates of $H_0$ at some fixed level $N$:
\bea
H_0\left\vert \psi \right\rangle \is
N\left\vert \psi \right\rangle\; 
\eea
Note that the level constraint $H_0 = N$ indeed eliminates one complex dimension:  it
fixes the absolute value of $Z^\alpha$ but also implements the $U(1)$ invariance under phase rotations $Z^I \to e^{i
\alpha} Z^I$.
The condition $H_0 = N$ plays the same role as the D-term constraint of the
usual gauged linear sigma model realization of $\mathbb{CP}^{3|4}$.

As anticipated, taking the quotient produces a finite dimensional Hilbert space, which we will denote by  $\mathcal{H}_{\mathbb{CP}^{3|4}}(N)$.  States of $\mathcal{H}_{\mathbb{CP}^{3}}
(N)$ are created by homogeneous degree $N$ polynomials in the creation operators, acting on the vacuum state.  Counting only the
four bosonic oscillators $Z_\alpha^\dag$, this represents a linear space of  dimension
\bea
\dim\mathcal{H}_{\mathbb{CP}^{3}}(N)\is \frac1 6 {(N+1)(N+2)(N+3)}\equiv k_{N}.
\eea
Taking into account the fermionic  oscillators, the dimension
of $\mathcal{H}_{\mathbb{CP}^{3|4}}(N)$ is:
\bea
\dim\mathcal{H}_{\mathbb{CP}^{3|4}}(N)\is k_{N}+4k_{N-1}+6k_{N-2}+4k_{N-3}%
+k_{N-4}=\frac{8}{3}N\left(  N^{2}+2\right) \equiv K_{N}.
\eea
We can now specify the form of the matrices $Z^\alpha$, by identifying them with the matrix elements of the
corresponding oscillators between states at some finite level $N$. Note, however, that $Z^\alpha$ does not commute with $H_0$ but  reduces
the level $N$ by one. The $Z^\alpha$'s thus define maps from $\mathcal{H}_{\mathbb{CP}^{3}}(N+1)$ to $\mathcal{H}_{\mathbb{CP}^
{3}}(N)$. In other words, the $Z^\alpha$ are non-square bosonic $k_{N}\times k_{N+1}$ matrices. In the supersymmetric case we view
the $Z$'s as $K_N \times K_{N+1}$ matrices. To write the gaussian matrix action (\ref{mmaction}),
we therefore need to define the matrix variables $\qQ$ and $\widetilde{\qQ}$ as non-square matrices of size
\bea
\label{qmatrix}
\qQ, \tQ\!\!   & \in & \! {\rm Mat}(k_{N+1}\! \times k_{N}).
\eea
More precisely, this is size of the the lowest superfield component of the matrix variables,
when acting on the lowest superfield component  of the Hilbert space. Since $\psi^i$ and $\psi^\dag_i$ carry
homogeneity charge $-1$ and $1$, higher superfield components have shifted ranks relative to the lowest
components. The matrices $Q$ and $\tQ$ then fill out $K_{N+1} \times K_{N}$ matrices. In what follows
we shall often leave the extension to the supersymmetric case implicit, so we write all expressions in terms of the bosonic matrix model:
\begin{equation} \label{boseGMM}
S_{\text{MM}} = \Tr\bigl(I_{\alpha \beta} \tQ Z^{\alpha} \qQ Z^{\beta}\bigr)
\end{equation}
The variables $\qQ$ and $\tQ$ are for the rest arbitrary matrices. A convenient
representation of the space of arbitrary $k_{N+1}\times k_{N}$ matrices is as the space of homogeneous
polynomials in $Z^\alpha$ and $Z^\dag_\beta$ of degree $1$, that is, polynomials in which each term contains one more
creation operator than annihilation operator,
with the relation that $Z^{N+1} = 0$ -- since any state in ${\cal H}_{\mathbb{CP}^3}(N)$
is mapped to $0$ after acting $N+1$ times with the $Z$'s.
Apart from this restriction, or after taking the large $N$ limit, we can thus view the matrix variables $\qQ$ and $\tQ$
as arbitrary sections of the degree $-1$ line bundle ${\cal O}(-1)$ defined on  $\mathbb{CP}^{3|4}$. At finite $N$, they are
sections of ${\cal O}(-1)$ defined on fuzzy $\mathbb{CP}^{3|4}$.

Finally, let us make a specific choice for the pairing ${\cal I}_{IJ}$. To this end, we decompose
the four coordinates $Z^\alpha$ into two two-component variables $\omega^{\dot a}$ and $\pi_a$  as
\be
Z^\alpha = (\spc \omega^{\dot a},\pi_a), \quad
\qquad
{Z}_\alpha = \left(\!\! \begin{array}{c}   \omega_{\dot a}\\
 \pi^a\end{array}\!\!\!\right)\, ,
 \ee
so that $Z_{\alpha} = I_{\alpha \beta} Z^{\beta}$, with $I^{\alpha \beta}$ the bosonic infinity bi-twistor.
In this notation,  we choose the matrix ${\cal I}_{IJ}$ to be of the following form
\bea
\label{inftwist}
\mathcal{I}_{IJ}=\left(\!
\begin{array}
[c]{ccc}						\spc \varepsilon_{\dot{a}\dot{b}} & 0 & 0 \\
0 &
\varepsilon^{ab} & 0 \\
0 & 0 & \eta_{ij}\! %
\end{array}
\right)
\eea
where $\eta_{ij}$ is a four index symmetric tensor.
It defines a pairing on $\mathbb{C}^{4|4}$
\be
\langle {Z}_1 {Z}_2 \rangle = \langle \pi_1 \pi_2 \rangle +
						   [ \omega_1 \omega_2]\,  + (\psi_1 \psi_2) .
\ee
where we introduced the spinor inner products
\bea
\label{spinpair}
\langle \pi_1 \pi_2 \rangle\!\spc = \! \spc  \varepsilon^{ab}\pi_{1a} \pi_{2b}\quad ; \quad
[ \smpc \omega_1\smpc \omega_2] \!\spc = \! \spc
\varepsilon_{\dot a \dot b} \omega_1^{\dot a} \omega_2^{\dot b}.
 \quad ; \quad(\psi_1 \psi_1)  \is \eta_{ij} \psi_1^i \psi_2^j.
\eea
In the following, we will freely raise and lower the spinor indices with the help of the corresponding $\varepsilon$
symbol. The canonical
commutation relations in the two-component spinor notation read\footnote{There is an unfortunate clash of notation for the
square brackets. The square brackets $[....]$ around two spinors without a comma in the middle denotes the anti-symmetric
pairing of two left handed spinors, where the brackets with a comma in the middle denote the usual commutator. As
another warning to the reader: the notation for hermitian conjugation here contains a raising operation for the indices:
the hermitian conjugate of $\pi_a$ and $\omega_{\dot a}$ is in fact not
equal to $\pi_a^\dag$ and $\omega_{\dot a}^\dag$, but rather
\be
(\pi_a)^\dag = \varepsilon^{a b} \pi_{b}^{\dag}, \qquad \quad (\omega_{\dot a})^\dag = \varepsilon^{\dot a \dot b}
\omega^\dag_b
\ee }
\bea
[\pi_a, \pi{^\dag}_{b}] = \varepsilon_{ab} \quad & ; & \quad [\omega_{\dot a}, \omega^{\dag}_{\dot b}] = \varepsilon_{\dot a
\dot b}.
\eea
The eigenvalue condition on the homogeneity operator, that defines the Hilbert space ${\cal H}_{\mathbb{CP}^3}(N)
$, takes the form
\be
H_0 |\Psi\ra = \Bigl(\pi^\dag_a \pi^a + \omega^\dag_{\dot a} \omega^{\dot a} + \psi_i^\dag \psi^i\Bigr) |\Psi\ra = N |
\Psi\ra\, .
\ee
Since $H_0$ keeps track of the spinor helicity, it is sometimes also called the helicity operator.
We see that states $|\Psi\ra$ in ${\cal H}_{\mathbb{CP}^3}(N)$ possess a large net helicity equal to $N$.

In the following, we will study the correlation functions of special bi-linear `current' operators computed in the
supersymmetric
matrix model at level $N$. We will focus on the leading behavior in the limit of large $N$.  In this limit, the size of the
Planck cells,
{\it i.e.} the scale of non-commutativity, tends to zero relative to the
total size of the projective space $\mathbb{CP}^{3|4}$. We can thus expect
that the large $N$ matrix model shares properties with some continuum field theory. In the naive continuum limit
of the matrix model action (\ref{boseGMM}), the trace over  ${\cal H}_{\mathbb{CP}^3}(N)$ turns into  an integral over
commutative twistor space $\mathbb{CP}^{3}$. The resulting free field action takes the form $S= \int_{\mathbb{CP}^{3}}
\tQ \Dbar Q$ with bosonic kinetic operator
\be
\label{contkin}
\Dbar = I_{\alpha\beta} Z^\alpha \frac{\partial}{\partial \overline{Z}_\beta} = \pi^a\frac\partial{\partial \overline{\pi}^a} +
\omega_{\dot a}\frac\partial{\partial \overline{\omega}_{\dot a}} .
\ee
At a heuristic level, this should be viewed as defining a kinetic term for chiral modes living in the fiber $\mathbb{CP}^{1|0}$ directions of the twistor space. Upon
summing over all points in the base space of the fibration $\mathbb{CP}^{1} \rightarrow \mathbb{CP}^{3} \rightarrow S^4$, this produces an action on the full twistor space.
For further details on the precise match to a continuum limit theory, we refer the interested reader to section 6 of \cite{TMMII}. We should point out, however, that the matrix theory is in fact better defined than the continuum field theory with the kinetic operator (\ref{contkin}). As we will see, the continuum theory is ultra-local in the sense that the modes only propagate in one direction, and stay localized on the other directions. Ultralocal theories do not really exist as continuum theories, since they typically lead to amplitudes that contain factors proportional to $\delta(0)$, the Dirac delta-function evaluated at 0. In the matrix theory, this divergence is automatically regularized. As we will see, this means that the large $N$ limit of the matrix model has to be taken with some care,  so as to preserve the UV
regulator~scale.

Of course, our motivation for studying the large $N$ limit of the matrix model (\ref{mmaction}) is not to regulate some unusual looking ultra-local theory on a complex 3-dimensional projective space. Projective superspace $\mathbb{CP}^{3|4}$ is the ${\cal N}=4$
supersymmetric  version of twistor space.
The matrix variables $\qQ$ and $\widetilde{\qQ}$ can thus be viewed as sections of bundles
on fuzzy twistor space, and via the twistor correspondence, they will then acquire a space-time
interpretation.

The twistor correspondence is based on the observation that, given a two component
spinor $\pi_a$ and a space-time point $x^{a\dot {a}}$ on complexified
Minkowski space, one can define a corresponding
two component complex spinor $\omega^{\dot a}$ via
$\omega^{\dot a}=ix^{\dot a a}\pi_{a}.$
This relation is invariant under simultaneous complex rescaling  of the
spinors $\pi_a$ and $\omega^{\dot a}$. For a given point $x$, it defines a  $\mathbb{CP}^{1}$,
called the {\em twistor line} associated with $x$. Similarly, a
point $(x^{\dot{a}a},\theta^{ia})$ in (chiral) Minkowski superspace specifies a bosonic $\mathbb{CP}^{1|0}$
in supertwistor space $\mathbb{CP}^{3|4}$, via
\bea
\omega^{\dot a}=ix^{\dot a a}\pi_{a}, \quad & & \quad
\psi^{i}=\theta%
^{ia}\pi_{a}. \label{bosonicline}%
\eea

Based on this space-time correspondence, we may thus  expect that a suitable class of correlation functions of the
large $N$
matrix model take on the form of space-time scattering amplitudes. In the following, we will consider two situations.
In the first case, we prescribe that the variables $\qQ$ and $\tilde{\qQ}$, in addition to being
$k_{N+1} \times k_{N}$ matrices,
also carry an index that transforms under the fundamental representation of an internal symmetry group, $U(N_c)$.
In the strict large $N$ limit, the current correlation function then reproduce the MHV amplitudes of ${\cal N}=4$
SYM theory with gauge group $U(N_c)$. Secondly, we will study a specially tuned
large $N$ scaling limit, where we simultaneously zoom in on a small region within the projective superspace,
in such a way that the scale of non-commutativity is kept fixed. The correlation functions in the resulting double
scaled matrix theory reproduce the MHV amplitudes of gravity, where the short distance cutoff coincides with the
Planck scale.

\section{Symmetries and Currents} \label{sec:SYMMS}

In this section we will take a first look at $\overline{D}$ the matrix model kinetic operator which acts via:
\bea
\label{kinetic}
\Dbar Q=I_{\alpha\beta}Z^{\alpha}Q Z^{\beta}\text{.}%
\eea
Since each $Z^\alpha$ has homogeneity one, and thus changes the level $N$
by one, $\Dbar$ defines a linear map between two spaces of matrices:%
\begin{equation}
\Dbar :{\rm Mat}(k_{N+1}\times k_{N})\rightarrow {\rm Mat}(k_{N}\times k_{N+1}).
\end{equation}
We see that the support and image space have the same dimension. Thus we should expect
the $\Dbar$ operator to be invertible as long as the anti-symmetric form $I_{\alpha\beta}$ is invertible.
We will investigate the inverse of $\Dbar$ later on.

In this section, we begin with a study of the symmetries of $\Dbar$. As we will see, this symmetry group is very large.
Via the analogue of Noether's theorem, this implies that the matrix model contains a rich collection of
current operators. We then study some preliminary aspects of correlators built from the symmetry currents of the
theory. These generate a $u(N_c) \times gl(k_N)$ current algebra. In the later sections, we show that
correlators of suitably defined currents compute scattering amplitudes.

\subsection{Global Symmetries}

To frame the discussion, we will look at the symmetries of the matrix model
through the lens of the twistor correspondence (\ref{bosonicline}). Hence we will view the
$Z^\alpha$'s as providing a twistor parametrization of space time. The symmetry transformations
then acquire the interpretation as space-time conformal transformations.

The main benefit of the twistor parametrization of space-time is that the conformal group is generated by linear
vector fields. Even in the usual discussions of twistor space, it is
standard to introduce canonically dual twistor variables $\widetilde{Z}_\alpha$ for a dual twistor space $\mathbb{P T}_{\bullet}$,
with  $[Z^{\alpha},\widetilde{Z}_{\beta}]=\hbar \delta_{\beta}^{\alpha}$,
and write the symmetry  generators as   ${\cal M}_{\alpha}{}^\beta = \tilde{Z}_{\alpha} Z^\beta$. Via the commutators,
these manifestly generate
$gl(4,\mathbb{C})$, which contains the 4D complexified conformal algebra $sl(4,\mathbb{C})$.
A choice of space-time signature amounts to picking an appropriate reality condition. For Euclidean signature,
one imposes the reality requirement $Z^\dag_\alpha = \tilde{Z}_\alpha$. This naturally leads to the commutation relation
(\ref{zcom}) and the construction of the finite dimensional Hilbert spaces ${\cal H}_{\mathbb{CP}^3}(N)$.

The 15 conformal generators
act on ${\cal H}_{\mathbb{CP}^3}(N)$ via the traceless $4\times 4$ matrix of operators
\bea
\label{mab}
\mathcal{M}_{\alpha\beta}\is Z_{\alpha}^{\dag}Z_{\beta}
\eea
where we lower indices via $I_{\alpha \beta}$ so that $Z_{\alpha} = I_{\alpha \beta} Z^{\beta}$.
We can associate to each conformal generator a linear operator that acts on the space of functions $\Phi$ on the
non-commutative twistor space (that is, on the space of linear operators $\Phi$ acting
on the Hilbert space ${\cal H}_{\mathbb{CP}^3}(N)$) via
\bea
{\cal M}^\circ_{\alpha \beta}  \Phi = {\cal M}_{\alpha\beta} \Phi - \Phi {\cal M}_{\beta\alpha}
\eea
A simple calculation shows that these operators all commute with the action of $\Dbar$ on $\Phi$:
\bea
 \bigl[\Dbar, {\cal M}^\circ_{\alpha\beta}\bigr] = 0.
 \eea
Based on this equation, it looks as if the kinetic operator $\Dbar$ of the matrix model preserves the full conformal
invariance. However, the matrix model action also involves a trace over the Hilbert space ${\cal H}_{\mathbb{CP}^3}(N)$ .
In order to be a true symmetry of the action, a charge needs to be hermitian with respect to the inner product,
that is used in defining the action.

Hermitian conjugation provides a reality condition, leaving us with the generators $u(4) \subset gl(4,\mathbb{C})$.
This is further broken to $SO(5)$ by the introduction of the anti-symmetric
bitwistor $I_{\alpha\beta}$.\footnote{The bitwistor transforms as a  6-component vector under $so(6) \simeq su(4)$.}
The hermitian charges that leave the bitwistor invariant are
${\cal M}_{(\alpha\beta)} = \frac 1 2\bigl( {\cal M}_{\alpha\beta} + {\cal M}_{\beta\alpha}\bigr)$,  which are the
10 generators of $SO(5)$. So these are the true global symmetries of the matrix model.  As we will see,  from the
space-time perspective,
the matrix model indeed naturally lives on the four sphere $S^4$.

The space-time interpretation becomes more evident when we write the symmetry generators in terms of
the two component spinors $\pi_a$ and $\omega^{\dot a}$.
We can then distinguish translations, conformal boosts, Lorentz rotations and the dilatation generator
\bea
\qquad P_{\dot a a}  =\omega^\dag_{\dot a}\pi_{a}%
\  \ & ; & \ \
J_{\dot a \dot b}   =\omega^\dag_{(\dot a }\omega_{\dot b)} %
\,,\nonumber\\[-3mm]
& & \hspace{3.5cm} D  ={\frac{1}{2}}\bigl(\omega^\dag_{\dot a}\omega^{\dot a}
-\pi^\dag_{a}\pi^{a}\bigr)
\\[-3mm]
K_{a \dot a}=\pi^\dag_{a}\omega_{\dot a} \ \ & ; & \ \ \widetilde
{J}_{ab}=\pi^\dag_{(a}\pi_{b)}\,,\nonumber
\eea
Hermitian conjugation leaves the $SU(2)\times SU(2)$ generators $J$ and $\tilde{J}$ intact, but acts on the other
symmetry generators via
\bea
P^\dag_{\dot a a} = K^{\dot a a} \quad & ; & \quad D^\dag = -D
\eea
The hermitian charges that also leave the anti-symmetric pairing (\ref{inftwist}) intact, are
the $SO(5)$ generators, $J$, $\tilde{J}$ and an additional four hermitian generators
given by:
\be \label{TRANSLATION}
\mathcal{P}_{\dot a a} = P_{\dot a a} + K^{\dot a a}\,
\ee
Thus, only the $SO(5)$ subgroup of the conformal group is unitarily realized. Adding supersymmetry is easy.
The supersymmetric kinetic operator $\overline{\mathcal{D}} Q =\, {\cal I}_{I\!\spc J} {\cal Z}^I Q {\cal Z}^J$
commutes with the generators of the complexified superconformal algebra $psl(4|4)$.
The supersymmetric hermitian charges that leave $\overline{\mathcal{D}}$ invariant generate the isometries of the
supersymmetric four sphere $S^{4|8}$.

The symmetry group of the kinetic operator $\Dbar$ is in fact much bigger than the global isometry group $SO(5)$.
Namely, we can consider operators that (similar to $\Dbar$) act on operators $\Phi$ with oscillators
from the left and from the right.  This allows us to define another class of symmetry generators
in the form of linear operators ${X}_{\alpha\beta}$  that act on matrices $\Phi$ via\footnote{There
also exists a
complex conjugate set of operators $X^*_{\alpha\beta} \Phi = Z^\dag_{[\alpha}\Phi Z_{\beta]}$, which also commute
with $\Dbar$.
These complex conjugate fields are less relevant for our later discussion.}
\bea
\label{hatex}
{X}_{\alpha\beta} \cdot \Phi \is Z_{[\alpha}\Phi Z^\dag_{\beta]}\, .
\eea
One easily verifies that $\Dbar$ also commutes with $X_{\alpha\beta}$
\bea
\label{xcomd}
\bigl[\spc \Dbar\! , {X}_{\alpha\beta} \spc \bigr] \is 0.\ \
\eea
The operators $X_{\alpha\beta}$, and the fact that they commute with $\Dbar$, will play an important role in what
follows. For reasons that will become apparent shortly, we will call ${X}^{\alpha\beta}$ position operators.
Equation (\ref{xcomd}) shows that $\Dbar$ is ultra-local, in the sense that it does not shift the value of the position
operators.  This gives us a first precise hint that up to a quantifiable amount of Heisenberg uncertainty at small
scales, the large $N$ matrix model preserves a notion of space-time locality.

\subsection{Currents}

In the accompanying paper \cite{TMMII},  we argue that the matrix model appears as part of a larger
theory that also contains a  gauge field $\bA$ that lives on fuzzy $\mathbb{CP}^{3|4}$.
The fields $\bA$ are linear operators which act on the Hilbert space ${\cal H}_{\mathbb{CP}^{3|4}}(N)$.
Here we will just focus on the subsystem obtained by setting the non-commutative gauge field $\bA$
to some fixed background value, appropriate to the description of scattering states.

From the perspective of the gaussian matrix model, this coupling is obtained by introducing a $U(N_c)$ flavor symmetry under which the
$\qQ$ transforms in the fundamental and $\tQ$ in the anti-fundamental. In other words each variable $Q$ and $\tQ$ defines an $N_c$-component vector
of $k_{N+1} \times k_{N}$ matrices. Gauging this symmetry results in the action:
\bea
\label{gaugact}
S \is  
{\rm Tr}\bigl(\tQ \spc \Dbar_{\bA} \qQ \bigr)
\eea
where the covariant derivative $\Dbar_{\! \spc \bA} = \Dbar + \bA$ acts on fields $\qQ$ via
\bea\label{covderiv}
\Dbar_{\mathcal{A}}\spc  Q \is I_{\alpha\beta} (Z^\alpha + \bA^\alpha) \qQ \spc Z^\beta
\eea
The $U(N_c)$ gauge field $\bA$ represents some fixed $(0,1)$ form on the non-commutative twistor space.
The action (\ref{gaugact}) is invariant under gauge transformations
\bea
\label{trafo}
\delta_f \qQ= f \qQ,	\qquad \delta_f \tQ= -\tQ f,	\qquad \delta_f \bA_\alpha =[Z_\alpha+\bA_\alpha,f]
\eea
where $f= f_\aA(Z,Z^\dag) \tau^\aA$,
with $\tau^\aA_{ij}$ an element of the Lie algebra of $U(N_{c})$,  denotes an infinitesimal gauge variation.
A basic observation, which will have important consequences later, is that  even when the color gauge group $U
(N_c)$ is abelian,
$N_c=1$, the gauge transformations retain their non-abelian character. Indeed, a $U(1)$
gauge theory on a non-commutative space is still non-abelian. As explained in \cite{TMMII}, the
matrix model enjoys a more general $gl(k_{N})$ symmetry, which acts by left multiplication on the $Q$'s and
right multiplication on the $\tQ$'s. This is automatically gauged due to the presence of the bulk gauge field. So we see
that there is actually a $u(N_c) \times gl(k_{N})$ gauge symmetry.

Fixing the background value of $\mathcal{A}$, the coupling between the matrix
current and the bulk gauge field provides a class of vertex operators for the theory:
\begin{equation}
\mathcal{J}(\mathcal{A}) = \Tr (I_{\alpha \beta} \mathcal{A}^{\alpha} Q Z^{\beta} \tQ).
\end{equation}
The computation of current correlators then reduces to specifying
a choice for the background field $\mathcal{A}$.

The possible background gauge fields are dictated by the equation of
motion $\mathcal{F}_{(0,2)} = J_{(0,2)}$ which relates
the $(0,2)$ component of the curvature for $\mathcal{A}$ to
a choice of background source. Working to linearized order
in the bulk equations of motion, we see that there is a
special class of solutions of the form:
\bea
\label{linear}
\bA^\alpha \is Z^\alpha V, \qquad \quad  V = V_\aA \otimes \tau^\aA
\eea
where here $\tau^{\aA}$ is an element of the Lie algebra $u(N_c) \times gl(k_N)$
and for now, $V_\aA$ is some arbitrary function of $Z^\alpha$ and $Z^\dag_\alpha$.
This choice is an accord with the fact that if we had provided a source for
the gauge field by activating a vev for $Q$ and $\tQ$, the gauge symmetry
would have been broken. The zero energy configuration would then have been a
gauge field of the form $\bA^{\alpha} = Z^{\alpha} V_R - V_L Z^{\alpha}$. Using the
residual generators of $gl(k_N)$ not contained in $u(k_N)$, this can be put
in the form of equation \eqref{linear}.

Plugging the linearized gauge field (\ref{linear}) back
into the matrix model action (\ref{gaugact}), the
corresponding moment of the current reads
\begin{equation}
\label{jveen}
{\cal J}(V) =  \Tr \Bigl(I_{\alpha \beta}\spc Z^\alpha \tQ \spc Z^\beta V  {\qQ} \spc
\Bigr)
\end{equation}
In this expression, we recognize the kinetic operator $\Dbar$ that appears in the free action. So we can adopt a
more compact notation, and write
\bea
\label{jvee}
{\cal J}(V) =\spc \spc {\rm Tr} \bigl(V {\qQ}\spc \Dbar{\tQ} \bigr).
\eea
In the subsequent sections, we will be interested in computing the correlation functions
of a product of several of these currents in the matrix model.

\subsection{Current Algebra}

Due to the appearance of the kinetic operator in
the definition  (\ref{jvee}), the currents ${\cal J}(V)$ vanish on shell,
${\it i.e.}$ whenever $\Dbar \tQ = 0$. As a result, if we
study the correlation function of a number of current operators, a given
current ${\cal J}(V_i)$ can be non-vanishing only at the location of
other current insertions. From the form (\ref{jvee}) of the current,
we see that ${\cal J}(V)$ represents the effect of
performing ``half'' a gauge transformation
\bea
\label{vact}
\delta_V \qQ = V \qQ  \quad & ; & \quad \delta_V \tQ = 0.
\eea
So to compute current correlation functions, all we need to do is to perform this substitution in the functional integral.

To make this explicit, consider two neighboring currents. Since the $Q$ and $\tQ$ are just free fields,
we can perform a single Wick contraction
\bea
{\rm Tr} \Bigl( V_1 \qQ \spc\Dbar  \underbracket[.5pt]{{\! \tQ} \spc \Bigr)\; {\rm Tr} \Bigl( {\qQ}\!\!}\spc\;(\!\spc \Dbar
{\tQ}) \spc V_2\Bigr) =\,
{\rm Tr} \left( V_1  {\qQ} \spc (\!\spc \Dbar
{\tQ}) \spc V_2 \right)
\eea
The right-hand side again looks like a current. So we derive the operator product relation
\bea \label{gaugealg}
\underbracket[.5pt]{\!\!\! {\cal J}(V_1) \, {\cal J}} (V_2) \is \mathcal{J}(V_2 V_1)
\eea
where the underbracket denotes a single Wick contraction. In analogy with continuum field theory,
we can thus define a commutator algebra of currents by subtracting the two ways of performing the
Wick contraction
\bea
\bigl[ {\cal J}(V_1), {\cal J}(V_2)\bigr] = {\cal J}\bigl([V_2,V_1]\bigr)
\eea
The current algebra is isomorphic to the Lie algebra $u(N_{c}) \times gl(k_N)$
of the $V$ matrices.

The computation of the correlation functions of currents thus completely trivializes,
especially when we take the large $N$ limit. In this case, we can perform the
successive single Wick contractions to find
\bea
\Bigl\la {\cal J}(V_1) \ldots {\cal J}(V_n) \Bigr\ra
\is \tr\bigl(V_n \ldots V_2 \spc V_1\bigr) + {\rm permutations}
\eea
where the sum is over all orderings of the symmetry generators as well as multi-trace contributions.
So all the physics goes into determining the natural set of $V$ generators that we should consider.

As the reader will have noticed, even when the color gauge group is  $U(1)$, the current algebra remains non-commutative. Indeed, the $U(1)$ acts not just via phase rotations, but also as a diffeomorphism on the fuzzy
twistor space. This is a first indication that the theory may contain a gravitational sector. To isolate the gravitational physics, it is natural to focus on states which in the commutative context would be neutral under the $U(1)$. As explained in \cite{TMMII}, this is accomplished by introducing a compensator gauge field $\widetilde{\mathcal{A}}^{\beta}$ and viewing the matrix fields
$Q$ and $\tQ$ as bifundamentals under a non-commutative
$u(N_c) \times u(1)$ gauge symmetry. The covariant derivative of equation (\ref{covderiv}) is then replaced by
$\overline{D}_{\mathcal{A}, \widetilde{\mathcal{A}}} Q = I_{\alpha \beta} (Z^{\alpha} + \mathcal{A}^{\alpha}) Q (Z^{\beta} + \widetilde{\mathcal{A}}^{\beta})$. The gravitational gauge symmetry then corresponds to the linear combination of $u(1)$'s which acts on the defects via the adjoint action. Note that in the commutative context, the mode would have been neutral under this adjoint $u(1)$ action. In our context, this acts via pure diffeomorphisms.

Let us next consider the corresponding gravitational currents. There are two $gl(k_N)$ symmetry currents, given by $\mathcal{A}^{\alpha} = Z^{\alpha} V$ and $\widetilde{\mathcal{A}}^{\beta} = \widetilde{V} Z^{\beta}$, where $V$ and $\widetilde{V}$ are functions of $Z$ and $Z^{\dag}$. The adjoint $u(1)$ action corresponds to setting $V =- \widetilde{V}$. This leads to an additional set of gravitational currents:
\begin{equation}\label{gravcurr}
\mathcal{T}(V) = \Tr \Bigl( \bigl[V , Q\bigr] \overline{D} \tQ \Bigr)
\end{equation}
The analogue of equation (\ref{gaugealg}) is then:
\bea \label{gaugealg}
\underbracket[.5pt]{\!\!\! {\cal T}(V_1) \, {\cal T}} (V_2) \is \mathcal{T}\bigl([ V_2 ,  V_1 ]\bigr)
\eea
In this case, the $V$'s directly act via commutators. The commutator algebra of the $V$'s should be
viewed in the commutative limit as the algebra of vector fields on twistor space. We will make this more precise
when we turn to the computation of scattering amplitudes. To this end, we now turn
to the space-time interpretation of the matrix model.

\section{Space-Time}\label{sec:spacetime}

Having shown that the gaussian matrix model enjoys a number of symmetries,
in this section we turn to their 4D space-time interpretation. In particular, we determine a
fuzzy twistor correspondence between points of a 4D space-time and fuzzy $\mathbb{CP}^{1}$'s. Using this
interpretation, we can view the matrix variables $\qQ$ and $\widetilde{\qQ}$ as fields on non-commutative
twistor space. Via the twistor correspondence, they will then acquire a space-time interpretation.

\subsection{Twistor Lines}

Given the appearance of the symmetry algebra $SO(5)$, we
should expect some connection with space-time physics. Here we develop the notion of
a ``coherent state'' $| x , \lambda)$ which is associated with a spacetime point $x^{\dot a a}$ and a local
coordinate $\lambda$ on a $\mathbb{CP}^{1}$. The extension to the supersymmetric situation will be straightforward, and is discussed in
\cite{TMM, TMMII}. To begin, we start with a normalized state $|0 , 0)$ which up to an overall normalization is the unique state
annihilated by the oscillators $\omega^{\dot a}$ and $\pi_{2}$. Acting by $SO(5)$ generators, we can sweep out the rest
of $\mathcal{H}_{\mathbb{PT}}$. The states of a fuzzy $\mathbb{CP}^{1}$ are obtained by acting with
$\widetilde{J}$, the $su(2)$ subalgebra built from just the $\pi$ oscillators. We refer
to a holomorphic point on this $\mathbb{CP}^{1}$
as a state $\vert 0 , \lambda )$ which satisfies:
\begin{equation}
\epsilon^{ab}\lambda_a \pi_b \, |0 , \lambda ) = 0 \,\,\,\text{and}\,\,\, \omega^{\dot a} \spc |0, \lambda) = 0\,  .
\end{equation}
where $(\lambda_1 , \lambda_2)$ are homogeneous coordinates of the
commutative $\mathbb{CP}^{1}$ and $\lambda = \lambda_{2} / \lambda_{1}$ is an
affine coordinate. The space of all $\vert 0 , \lambda)$'s
are mapped to each other via the $SO(4)$ generators $J$
and $\widetilde{J}$. The equivalence class of all such states is then
a fixed point of $SO(4)$, corresponding to the south pole of an $S^{4}$.

Starting from the south pole of the $S^{4}$, we can now sweep out the remaining states by
$SO(5)$ generators. Acting via
$x \cdot \mathcal{P} = x^{\dot a a} P_{\dot a a} + x_{\dot a a} K^{\dot a a}$
of equation (\ref{TRANSLATION}), we obtain states:
\begin{equation}
\vert x , \lambda ) = \exp(i x \cdot \mathcal{P}) \vert 0 , \lambda )
\end{equation}
In this way we build up a spin $N/2$ $su(2)$ bundle fibered over $S^{4}$. Note that the
transformations $\exp(i x \cdot \mathcal{P})$ are unitary, and do not alter the norms of states.

The flat space limit corresponds to the Wigner-In\"on\"u contraction of $SO(5)$ where we
rescale the generator $P_{\dot a a}$ relative to $K^{\dot a a}$. In this limit $\mathcal{P} \rightarrow P$,
and the states $\vert x , \lambda )$ satisfy:
\begin{align}
\label{cohersup}
\epsilon^{ab}\lambda_a \pi_b \, |x, \lambda ) & = 0 \\
\label{cohertwo}
\bigl(\omega^{\dot a} - i x^{\dot aa}\spc \pi_a \bigl) |x, \lambda) & = 0
\end{align}
The second line is nothing but the usual twistor equation associated with a
space-time point $x^{\dot a a}$, but now interpreted as a holomorphic operator
equation. In other words, we recover the expected correspondence
between a point $x^{\dot a a}$ of complexified Minkowski space
and a (fuzzy) $\mathbb{CP}^{1}$.

By a similar token we can introduce bra states $( x , \lambda | = (0 , \lambda| \exp(- i x \cdot \mathcal{P}) $.
We provide the precise definition of $(0 , \lambda |$ in section \ref{sec:WARMUP}. In the flat space limit
we obtain bra states annihilated by $\omega^{\dag}_{\dot a } - i x_{\dot a a} \pi^{\dag a}$. Note that both the bra
and ket states can be extended to holomorphic $x^{\dot a a}$. This is an important feature of
twistor geometry which is preserved by the matrix geometry.

Having established a connection with classical twistors and the 4D continuum space-time, let us
now discuss some additional features of commutative twistors. See \cite{Penrose:1986ca, WardWells}
for additional review. In twistor theory \cite{Penrose:1967wn, Penrose:1968me},
the identification between space-time points and complex lines in
twistor space is a correspondence at the level of holomorphic geometry.
Complexified conformally compactified Minkowski space is given by the zero locus of
the Klein quadric in $\mathbb{CP}^{5}$
\be
\epsilon_{\alpha\beta\gamma\delta}X^{\alpha\beta}X^{\gamma\delta} = 0\ .
\label{KleinQ}
\ee
Here $X^{\alpha\beta}=-X^{\beta\alpha}$ is a four index anti-symmetric tensor,
defining the six homogeneous coordinates of $\mathbb{CP}^{5}$. The
constraint (\ref{KleinQ}) is automatically solved by introducing a pair of points in
twistor space, with homogeneous coordinates $U^{\alpha}$ and $V^{\beta}$, via%
\be
X^{\alpha\beta} = U^{[\alpha} V^{\beta]}\, .
\ee
Since two twistor points $U$ and $V$ determine a line $Z= a U+ bV$ in $
\mathbb{CP}^3$, one recovers the map between space time points
and twistor lines.

\def\ccdot{\! \cdot \! \spc}
\def\xX{\mbox{\fontsize{13.5}{14}\selectfont $x$}}

The  homogeneous coordinates  $X^{\alpha\beta}$ are sensitive  only to the conformal
structure of space-time.  Conformal symmetry is broken by designating a choice of two index
anti-symmetric bitwistor, called the infinity twistor, denoted by $I_{\alpha\beta}=-I_{\beta
\alpha}$.
The `inverse' bitwistor is
denoted by $I^{\alpha\beta}= \frac{1}{2} \varepsilon^{\alpha\beta\gamma\delta}%
I_{\gamma\delta}$.
The infinity twistor defines an anti-symmetric pairing
\bea
\la \spc Z\spc W \ra \is I_{\alpha\beta}Z^\alpha W^\beta
\eea
and allows us to raise and lower the index of the twistor coordinates $Z^{\alpha}$ via
$Z_\alpha \equiv I_{\alpha\beta}Z^\beta$. With
the help of the infinity twistor, we can define affine space time coordinates
\bea
\label{xrat}
 \xX_{\alpha\beta} = \frac{X_{\alpha\beta}}{X_0} , \qquad X_0 = I^{\alpha\beta} X_{\alpha\beta}
\eea
The matrix $\mathcal{I}_{IJ}$ that features in the gaussian matrix model action is the infinity twistor
of $S^{4|8}$, the supersymmetric four-sphere.

We have already encountered this formulation of the twistor correspondence in the non-commutative
setting, in the form of the operators\footnote{Here we temporarily put a $\hat{\ }$ on the coordinate operators $\hat{X}_
{\alpha\beta}$,
to distinguish them from ordinary c-number space-time coordinates.}
$
\hat{X}_{\alpha\beta} \ccdot \Phi = Z_{[\alpha}\Phi Z^\dag_{\beta]}
$
introduced in equation (\ref{hatex}).
We now see the space-time significance of these operators: they allow us to define the notion of operators $\Phi(x)$
that are localized at a given space-time point $x$. In analogy with (\ref{xrat}), we shall sometimes
refer to an eigenstate of $\hat{X}$ as a matrix $\Phi(x)$ which satisfies:
\bea
\label{eigenv}
\hat{X}^{\alpha\beta} \ccdot \Phi(x) = \xX^{\alpha\beta}\spc\hat{X}^0 \ccdot \Phi(x)	
\eea
where $\hat{X}_0 = I^{\alpha\beta} \hat{X}_{\alpha\beta}$ and where $\xX_{\alpha\beta}$ are c-numbers.
Alternatively, we could have defined local operators $\Phi(x)$ as operators that satisfy the space-time coherent state
conditions from the left, and the hermitian conjugate conditions from the right
 \bea
 \label{twisteen}
 \bigl(\omega^{\dot a}\! -\!\spc i x^{a \dot a} \pi_a\bigr)\Phi(x)\spc = 0 \spc & ; & \ \spc  \Phi(x)\bigl(\omega^{\dag \dot{a}}
\! - \!\spc i x^{\dot a a} \pi^\dag_a \bigr)	= 0
 \eea
It is not difficult to show that the two definitions (\ref{eigenv}) and (\ref{twisteen}) of operators $\Phi(x)$, that are
localized in space time, are equivalent.\footnote{The proof is the same for the left twistor line equation and its
conjugate, and goes as follows:
$$
0 = \epsilon_{\alpha\beta\gamma\delta} Z^\beta \hat{X}^{\gamma\delta} \Phi(x) = \bigl(\epsilon_{\alpha\beta\gamma
\delta} Z^\beta \spc \mbox{\normalsize{$x$}}^{\gamma\delta}\bigr) \hat{X}^0 \Phi(x)
$$
Since $\hat{X}^0$ commutes with $Z^\beta$, and is invertible, this implies that $(\epsilon_{\alpha\beta\gamma
\delta} Z^\beta \spc \mbox{\normalsize{$x$}}^{\gamma\delta}) \Phi(x)
 = 0 $}
The precise relation between the eigenvalues $\xX_{\alpha\beta}$ in (\ref{eigenv}) and the flat space-time coordinates
$x^{\dot a a}$ that appear in the twistor line equation (\ref{twisteen}) is given in Appendix A.

The commutator of $X_{\dot a a}$ with the $SO(5)$ rotation
generators ${\cal P}_{\dot a a} =  P_{\dot a a} + K^{\dot a a}$ reads
\bea
\left[ {X}_{\dot{a}a},{\cal P}_{\dot{b}b}\right]   \is \epsilon_{ab}\epsilon_{\dot a \dot b}
{X}_0.
\eea
Near the south pole region, where $X_0$ is maximal, we can approximate $X_0$ by its maximal eigenvalue.
After rescaling  ${X}_{\dot{a}a}$ to $\xX_{\dot a a} = X_{\dot{a}a}/X_0$, this relation yields the Heisenberg commutation
relation between momenta and coordinates. 
Finally, since the position operators $X_{\alpha\beta}$ commute with the kinetic operator $\Dbar$
of the matrix model
\bea
[\Dbar\! , X_{\alpha\beta} ] \is 0.
\eea
Hence, the $\Dbar\! $ operator maps the local operators
$\Phi(x)$ (defined via the eigenvalue equation (\ref{eigenv}) or equivalently, the coherent state condition (\ref
{twisteen})) to another local operator $(\Dbar\Phi)(x)$
defined at the same space-time point $x$. The kinetic
operator ${\Dbar\, }$ thus acts along the twistor lines.

\subsection{Planck Scale}

We have seen that the non-commutative theory allows for the introduction of position operators
$X_{\alpha\beta}$ with a continuous spectrum of eigenstates. Of course, this does not mean that the
matrix model defines an exact local theory. Indeed, from the perspective
of the 4D space-time, the fuzzy twistor space corresponds to truncating the angular momentum on the
$S^{4}$. This limits the angular resolution of the 4D theory. The number
of independent spherical harmonics on $S^{4}$ at level $N$ is of order $N^{4} / 12$, with
corresponding resolution area $\ell^{2}_{pl} \sim \ellcc^2 / N$ \cite{TMMII}.

Another way to see the presence of this minimal length scale is by evaluating the
overlap of position eigenstates $(U| = (x,\lambda|$ and $|V) = |y,\xi)$, for different space-time points $x$~and~$y$:
\bea
\label{overl}
(U| V) \is (x,\lambda|y, \xi) = (0 , \lambda | e^{-i x \cdot \mathcal{P}} e^{i y \cdot \mathcal{P}} | 0 , \xi).
\eea
The composition $e^{-i x \cdot \mathcal{P}} e^{i y \cdot \mathcal{P}}$ is again an $SO(5)$ rotation.
At small displacements, it corresponds to a translation in the direction $r_{\dot a a} = (x - y)_{\dot a a}$. Next,
we can use the fact that any $SO(5)$ rotation operator can be factorized as a product $R_5 = (R_4) \spc  R_\theta \spc( R_4)$ where each $(R_4)$ factor is an $SO(4)$ rotation, and where $R_\theta$ is the special rotation matrix $R_\theta = e^{i \theta \hat{r} \cdot{\cal P}}$ with $\hat{r}_{a\dot a} $ a unit $2 \times 2 $ matrix proportional to $r_{\dot a a}$. The diagonal $SO(5)$ rotation $R_\theta$ can be thought of as the rotation that transports the point $x$ along a great circle to $y$. The rotation angle is the arc length
\be
\theta = {|x-y|}/\ellcc
\ee
$R_\theta$ acts on the four twistor coordinates via the simple rotation (see for example \cite{Holman:1969cm})
\be
R_\theta : \quad  \left(\!   \begin{array}{c}  \pi_1 \\[.5mm] \pi_2 \\[.5mm]\omega_{\dot 1} \\[.5mm]\omega_{\dot 2} \end
{array}\!   \right) \; \to\;
\left(\!   \begin{array}{c}  \cos\frac{\theta}{2}\, \pi_1 \! +\! i\sin\frac{\theta}{2} \, \omega_{\dot 1}\! \\[.5mm] \cos\frac{\theta}{2}\,
\pi_2 \! -\! i \sin\frac{\theta}{2} \, \omega_{\dot 2}\! \\[.5mm] \cos\frac{\theta}{2}\, \omega_{\dot 1} \! +\! i \sin\frac{\theta}{2} \,
\pi_{1}\! \\[.5mm]\cos\frac{\theta}{2}\, \omega_{\dot 2} \! -\! i\sin\frac{\theta}{2} \, \pi_2\! \end{array} \! \right)
\ee
Ignoring for now the $SO(4)$ part of the rotation, one can easily compute the matrix element
by letting this transformation act on the ket state $|0,\lambda)$. Using that both the bra
and ket state contain only $\pi$ oscillators, one immediately finds that the answer collapses to
\bea
\label{noverlap}
(0,\lambda| R_\theta | 0, \xi)  \is \Bigl(\cos\mbox{\large $\frac \theta 2$}\,\Bigr)^{N} (0,\lambda | 0, \xi)
\eea
This is the expected behavior of an $SO(3)$ transformation with
rotation angle $\theta$  acting on a spin $N/2$ representation.

We are interested in the leading behavior at large $N$ and small $\theta$:
\be
\label{gaussian}
\Bigl(\cos\mbox{\large $\frac{|x-y|} {2\ellcc}$}\,\Bigr)^{N} \to\; \; \exp\Bigl({- \frac{N |x-y|^2}{8 \ellcc^2}}\Bigr)\; \  \to \  \ell_{pl}^{4}
\delta^4(x-y)
\ee
Here we introduced the UV length scale $\ell_{pl}$ via
\be
\label{lplanck}
\ell_{pl}^2 = \frac{8 \pi \ellcc^2}{N}
\ee
The parameter $\ell_{pl}$ represents a short distance cutoff for our theory. In the last step in (\ref{gaussian}) we took the large $N$, large $\ellcc$ limit while keeping $\ell_{pl}$ very small but finite. Finally, we can verify that the $SO(4)$ part of the $SO(5)$ rotation indeed drops out, because (i) the states $|0,\xi)$ and $(0, \lambda|$ do not depend on the $\omega$ oscillators, and (ii) the delta function enforces that the rotation parameter $r_{\dot a a}$ vanishes anyhow.

\section{Chiral Field on $\mathbb{CP}^{1}$}\label{sec:WARMUP}

Having presented the gaussian matrix model and established that it retains a
natural 4D space-time interpretation, we would like to study the correspondence between
correlators of the matrix model and 4D physics. To this end, in this section
we compute the exact form of the propagator for a
chiral boson or fermion system on a fuzzy ${\mathbb{CP}}^1$.

On a commutative $\mathbb{CP}^1$, the chiral free field action takes the form
\be
\label{freechiral}
S = \int_{\mathbb{CP}^1} \!\!  \tilde{\phi}\, \delbar \phi\spc
\ee
The chiral fields $\phi$ and $\tilde{\phi}$ can either both be fermions or bosons.
In principle, they can carry
arbitrary half integer spin $s$ and $1-s$, respectively.
In the following we will mostly restrict to the spin half case $s=\frac 1 2$, so that
$Q$ and $\tilde{Q}$ are both sections of the degree $-1$ bundle ${\cal O}
(-1)$.
Using projective coordinates $(\pi_a,\bar{\pi}_b)$, $a=1,2$, on $\mathbb{CP}^1$,
the $\delbar$ operator reads
\be
\label{cdelb}
\delbar = 
\pi_a \frac{\partial}{\partial \bar{\pi}_ a}
\ee
In this section, we are interested in constructing the non-commutative version of
the propagator of the
chiral fields.
In other words, we will be looking for the analogue of the Green's function $\Delta
(\pi,\lambda)$ associated
with the $\delbar$ operator. Using the projective notation $\langle \ppi  \spc
\lambda \rangle \equiv
\epsilon^{ab} \lambda_a\ppi_b,$ for the difference between two points, our task is
to solve the equation
\be
\label{greend}
\delbar \Delta(\pi,\lambda) = \delta(\la \pi\spc \lambda \ra )
\ee
In the commutative theory, this is trivially solved via\footnote
{In affine coordinates $\xi= \ppi_2/\ppi_1$ and $\lambda =
\lambda_2/\lambda_1$, it takes the
more familiar form  $\Delta(\xi,\lambda) = \frac{1}{\xi-\lambda}$}
\be
 \Delta(\pi,\lambda)  =  \frac{1}{\langle \spc \ppi \spc \lambda\spc \rangle}.
\ee
As we will see, although finding the non-commutative analogue of this
expression takes a bit more work,
the end result will be almost as simple.

Non-commutative $\mathbb{CP}^1$ is described by oscillators $(\ppi_a,{\ppi}^
\dag_a)$, with $a=1,2$,
satisfying the canonical commutation relation $\bigl[\ppi_a,\ppi^\dag_b\bigr] = \epsilon_{ab}.$
The oscillators act on finite $N+1$ dimensional Hilbert spaces ${\cal H}_{\mathbb
{CP}^1}(N)$,  specified
by the level constraint
\be
\label{level}
\epsilon^{ab} \ppi^\dag_a \ppi_b |\psi\rangle = N |\psi\rangle.
\ee
As before, we can think of the Hilbert space
${\cal H}_{\mathbb{CP}^1}(N)$ as the space of points on the $\mathbb{CP}^1$.
The chiral fields $Q$ and $\tQ$ represent arbitrary homogenous
polynomials in the
creation and annihilation operators $\ppi_a$ and $\ppi_a^\dag$ of a given
degree specified by their spin.
In the spin 1/2 case, they are
taken to be homogeneous functions with one more $\ppi^\dag$ than $\ppi$.
Hence, the fields do not act
within the same finite Hilbert space:
$\phi$ and $\tilde \phi$ both act as linear maps from ${\cal H}_{\mathbb{CP}^1}
(N)$ to ${\cal H}_{\mathbb
{CP}^1}(N\! + \! 1)$, and can thus be viewed
as arbitrary $(N+2)\times(N+1)$ matrices.

The non-commutative version of the action (\ref{freechiral}) reads
\be
\label{freecpone}
S = {\rm Tr}\bigl(\tilde{\phi}\, \epsilon^{ab}\ppi_a\phi\spc \ppi_b  \bigr),
\ee
where the trace is taken over ${\cal H}_{\mathbb{CP}^1}(N\!+\! 1)$.
Note that the kinetic operator
\be
\label{delbar}
\delbar \equiv
\epsilon^{ab}\ppi_a{}_L \ppi_b{}_R
\ee
via the action of the commutator on the $\pi^\dag_c$ dependence of $\phi$, indeed defines a direct analogue of the
Dolbeault operator (\ref{cdelb}). See \cite{Grosse:1994ed, Grosse:1995jt, Dolan:2007uf} for further discussion on the form of the
fuzzy Dolbeault operator.

Given the action, we can start to compute correlation functions. For this we need
the explicit form of the
propagator $\Delta = 1/\overline{\partial}$.
Mathematically, the $\delbar$-operator  defines a linear map from the space of $
(N + 2) \times  (N + 1) $
matrices to the space of $(N+ 1) \times(N + 2) $  matrices.
Since the support and image have the same dimension, this map is expected to
be invertible.
This is indeed obvious from the oscillator representation: $\delbar$
has no zero modes, since $\phi$ always contains at least one $\ppi^\dag$
oscillator.
We can thus define the non-commutative version of the propagator as the inverse
of this linear map.

\subsection{Affine Coordinates}

While the commutative theory (\ref{freechiral}) enjoys full conformal invariance,
the non-commutative deformation breaks the conformal group to
the group of global $SU(2)$ rotations acting on the doublet of oscillators $\pi_a$.
The non-commutative $\mathbb{CP}^1$ is indeed equivalent to a
fuzzy two-sphere.  In the following, however, we will not use this
global $SU(2)$ perspective, because we wish to preserve the holomorphic properties of the theory as much as possible. To this
end, we will choose to work in a local affine patch with coordinate $\lambda = \lambda_2/\lambda_1$.
As we will see, this will allow a formulation in which conformal symmetry will
naturally re-emerge once we take the large $N$ limit.

\def\nN{{\!\smpc {}_N}}
\def\ccN{\mbox{\footnotesize $\cal N$\!}}

Let us introduce the following number basis of ${\cal H}_{\mathbb{CP}^1}(N)$
and its dual
\bea
\label{ndef}
 |\spc n \spc ) 
 \is \frac{
(\ppi_2^\dag)^{N-n} (\ppi_1^\dag)^{n} }{(N\! - \! n)!\spc n! } \spc
|\spc 0 \spc \rangle , \qquad \qquad
( n \spc | =  \la 0 |
{(\ppi_2)^{n}
(\ppi_1)^{N-n}}.
\eea
The normalization factors are convenient for our present discussion.
This basis is canonically normalized
\be
\label{inpro}
( n| m ) 
=   \delta_{n, m}\qquad
\ee
where $n$ and $m$ both run from $0$ to $N$. The Hilbert space ${\cal H}_
{\mathbb{CP}^1}(N)$
contains a continuous family of coherent states, labeled by points $\lambda$ on
the commutative $
\mathbb{CP}^1$, defined via
\bea
\label{lambdef}
|\spc \lambda \spc  ) =\theta_\nN(\lambda) \,\spc 
\sum_{n=0}^N \lambda^{n}  \spc |\spc n\spc ),
\eea
where $\theta_\nN(\lambda)$ is a normalization factor. A geometrically natural requirement is that
$\lambda$ is invariant under simultaneous transformation $\pi_1 \leftrightarrow \pi_2$ and $\lambda \leftrightarrow
\lambda^{-1}$. This leads to
\be
\label{thetadef}
\theta_\nN(\lambda) = \frac{1}{1 - \lambda^{N+1}}
\ee
In the large $N$ limit, this becomes a step function:
\be
\label{resultt}
\theta(\lambda) =  \left\{ \begin{array}{cc} \, 1 \qquad  \text{for}\  |\lambda| < 1 \\[3mm] 0
\qquad \text{for} \ |\lambda| > 1
\end{array} \right.
\ee
The states $|\lambda)$ satisfy the coherent state condition
\bea
\label{coherent}
(\ppi_2 - \lambda \ppi_1) |\spc \lambda \spc ) 
 = 0
\eea
which shows that $\lambda$ can be thought of as the classical value of the affine
coordinate $\lambda_2/\lambda_1$.
We may  write (\ref{coherent}) in a slightly more covariant notation as
\bea
\epsilon^{ab} \lambda_a \pi_b |\spc \lambda \spc ) 
 = 0
\eea
with $\lambda_a = (1,\lambda)$.
We can call the states $| \spc \lambda \spc )$ 
`position eigenstates', although there obviously does not exist any unitary
position operator of which they
are eigenstates.
The state $|0)$  corresponds to the position state at the origin $\lambda = 0$,
while the state $|N)$
corresponds to the point at infinity
\be
\label{north}
|N) = |\infty).
\ee
We will sometimes call $|0)$ the south pole state, and $|\infty)$ the north pole
state.

At this point it is useful to introduce the non-commutative notion of the affine
coordinate chart.
The main advantage is that the chiral fields will become square matrices.  In our
setting, specifying an
affine coordinate system amounts to picking a `canonical' embedding of
${\cal H}_{\mathbb{CP}^1}(N)$ inside of ${\cal H}_{\mathbb{CP}^1}(N +1)$, or
equivalently, a projection
from ${\cal H}_{\mathbb{CP}^1}(N+ 1)$ onto
${\cal H}_{\mathbb{CP}^1}(N)$. Choosing the coordinate $\lambda = \pi_2/\pi_1$
amounts to identifying the states in both spaces via the
action of the $\pi_1$ oscillator. In particular, position eigenstates are related via
\be
\label{iso}
|\lambda )_\nN = \pi_1 |\lambda)_{{\!}_{N+1}} .
\ee
This map projects out the north pole state (\ref{north}), since $\pi_1 |\infty) = 0$.
We will call this coordinate
chart the south pole patch.
The restricted Hilbert space, with the north pole state projected out,
will be denoted by ${\cal H}^\prime_{\mathbb{CP}^1}(N\! +\! 1)$. The map (\ref
{iso}) provides an
isomorphism
\be
{\cal H}_{\mathbb{CP}^1}(N) \simeq {\cal H}^\prime_{\mathbb{CP}^1}(N\! +\! 1)
\ee
The chiral free fields in the affine coordinate patch are defined as $\Phi =  \ppi_1
\phi$ and $\tilde \Phi =
\ppi_1 \tilde \phi$.
The redefined fields both act as linear maps from ${\cal H}_{{\mathbb CP}^1}(N)$ to itself, and
thus specify
square $(N+1) \times (N+1)$ matrices. We will use the isomorphism (\ref{iso})
repeatedly in what follows.

As one would expect, the dual Hilbert space is naturally viewed as describing the
opposite patch with affine coordinate $\xi = \pi_1/\pi_2$.
We will call this the north pole patch. In adhering to the usual notions of twistor theory,
we seek a suitable \textit{holomorphic} notion of a bra state.
At first sight, however (since bra states cannot be annihilated by a linear
combination of annihilation operators)  there is no obvious dual basis of position eigenstates,
which are annihilated by the holomorphic operator $\ppi_1 -\xi \ppi_2$.
We can still define coherent states $(\xi|$ via
\be
(\xi|\,  = \, \theta_\nN(\xi)\; \sum_{n=0}^N \spc \xi^n\spc (n|\spc,
\ee
with $\theta_\nN(\xi)$ defined in (\ref{thetadef}).
A straightforward calculation shows that the holomorphic coherent state condition is violated at the north and south
pole
\footnote{Here we use the action of $\pi_2$
to define a canonical embedding of the dual Hilbert space  ${\cal H}^*_{\mathbb{CP}^1}(N)$
inside ${\cal H}^*_{\mathbb{CP}^1}(N\! +\! 1)$.}
\be
\label{result}
(\xi| (\pi_1 - \xi \pi_2) \spc =\spc \theta_\nN(\xi)\spc (0| - \theta_\nN(\xi^{-1})\spc (\infty|
\ee
Here $(0|$ and $(\infty|=(N|$ denote the dual north and south pole state in ${\cal H}^*_{\mathbb{CP}^1}(N+1)$.
The dual north pole state $(0|$ is located at $\xi =0$ and the south pole state $(N| = (\infty|$  is  the
place where $\xi=\infty$. Both states will play a special role in what follows. Note that the
inner product pairs the dual north and south  pole states with their polar opposites
\be
(0| 0) =  (\infty | \infty ) = 1, \qquad (\infty| 0) = (0|\infty) = 0.
\ee
In this sense, our inner product is similar to the BPZ inner product in the radial
quantized formulation of 2D conformal field theory.

In the large $N$ limit, the factor $\theta_\nN(\xi)$ becomes  a step  function: it is equal to $1$ on
the northern hemisphere where $|\xi|< 1$, and vanishes on the southern hemisphere where $|\xi|>1$.
So after taking the large $N$ limit, eqn (\ref{result}) reduces to
\bea
\label{resulttt}
(\xi | (\pi_1 - \xi \pi_2) =  \left\{ \begin{array}{cc} \, (0| \qquad \  \text{for} \ |\xi|
< 1 \\[3mm](\infty| \qquad
\text{for} \ |\xi| > 1 \end{array} \right.
\eea
This is our desired intermediate result. It shows that the dual coherent state $
(\xi|$ are position eigenstates, modulo a source term localized at
the corresponding pole.

Let us compute the overlap between the position eigenstates. A direct
calculation shows that
\bea
\label{mulambd}
( \spc \xi \spc | \spc \lambda \spc )
= \frac{ \theta(\xi)  - \theta
(\lambda^{-1})}{1- \xi
\lambda}
\eea
This equation reveals, as expected, that $\xi$ and $\lambda$ are reciprocal
affine coordinates. Note,  however, that the pole in the denominator is spurious.
The step functions do not allow $\xi$ and $\lambda$ to be located on the same
hemisphere: whenever they do, the numerator vanishes.

The standard way to overcome this obstacle is via analytic continuation.
Consider the overlap $(\xi_1|\lambda_2)$ and let $\lambda_1$ be the reciprocal coordinate  to $\xi_1 =
\lambda_1^{-1}$. We wish to {\it define} the south patch state $(\lambda_1|$
via analytic continuation of the north patch state $(\xi_1|$. However, here we
meet a subtlety. At infinite $N$, the step functions are non-analytic at the equator,
while at {finite} $N$ the step functions $\theta_N(\xi)$  are perfectly analytic.
Our approach is to first take the large $N$ limit, and then analytically continue.
See Appendix B for more discussion of this issue.

The upshot is this: we {\it define} the state $(\lambda_1|$ such that its overlap
with $|\lambda_2)$ is given
by the analytic continuation of (\ref{mulambd}),
starting from the region $\theta(\xi) - \theta(\lambda^{-1}) = 1$. Hence in the
strict large $N$ limit, we have\footnote{In going from the patch near the north
pole to the patch near the south pole, we have used the fact that
the bra states transform as $1/2$-differentials, and so in passing from one patch to the other,
transform as $( \xi | \rightarrow \frac{1}{\lambda} (\lambda|$.}
\be
(\lambda_1 | \lambda_2) = \frac{1}{\lambda_1 - \lambda_2} \equiv \frac 1 {\la
\lambda_1 \lambda_2\ra}
\ee
Moreover, via the analytic continuation of (\ref{resulttt}), we learn that $(\lambda|$,
as defined this way, solves the ket state condition $( \lambda | \epsilon^{ab} \lambda_{a} \pi_{b} = (0|$
up to terms which vanish in the large $N$ limit.
This completes our construction of the state $(\lambda|$.

\subsection{$\mathbb{CP}^1$ Propagator}

The construction of  the propagator of the chiral free fields is now almost as
simple as in the commutative case, or possibly even simpler.
We first need to define the notion of a holomorphic delta function. Let $\lambda$
be a point on the commutative $\mathbb{CP}^1$, with associated coherent state $|\lambda)$.
Our proposed definition for the projective delta function localized at the point $
\lambda$ is as follows
\be
\label{projd}
\delta(\la \pi \lambda\ra) = |\lambda ) ( 0 | \, .
\ee
Let us motivate this definition.
Eqn (\ref{projd}) defines a projection onto the position eigenstate at $\lambda$,
which is similar to how a
commutative holomorphic delta function
acts on the space of functions. The right-hand side explicitly involves the special
state $(0|$ which
represents the point at infinity of the affine chart $\pi_1 = 1$.
The projective delta-function on the left-hand side seemingly does not depend on
such a choice -- but of
course it does once we choose an affine chart on $\mathbb{CP}^1$.

Given this definition of the projective delta function and the result (\ref{result}),
we now have a natural candidate for the propagator
\be
\label{cpoprop}
\Delta(\pi, \lambda) = |\lambda) (\lambda |
\ee
The verification is trivial:
\bea
\epsilon^{ab} \pi_a |\lambda) (\lambda|\pi_b \is  |\lambda) (\lambda|\epsilon^{ab}
\lambda_a\pi_b  = |
\lambda) (0|
\eea
up to corrections which are exponentially suppressed at large $N$. Given
the proposed identifications, this calculation provides the non-commutative
version of eqn (\ref{greend}) that defines the Green's function of the $\delbar$ operator.

\def\cmu{\tilde\mu^\vee}

\section{Scattering Amplitudes} \label{sec:SCATT}

Having studied correlators of the $\mathbb{CP}^{1}$ system, we now turn to correlators
of the full gaussian matrix model. In this section we propose a direct correspondence
between correlators of the matrix model and amplitudes of the 4D space-time theory.
This correspondence is defined in a double scaling limit where we
zoom in on a small neighborhood near the south pole of the $S^{4}$:
\be
\label{scaling}
N \to \infty, \ \ \ellcc \to \infty, \qquad \ell_{pl}^2 = \frac{ \ellcc^2}{N}\ \ {\rm fixed}
\ee
Scattering in the 4D theory proceeds as follows. We  prepare states ``at infinity'', corresponding to the boundary of
the small patch near the south pole. The rescaled patch defines our 4D spacetime for the
scattering experiment. To have a notion of lightlike momenta, we compute the values of the correlators in Euclidean signature, and then
analytically continue to lightlike values of the complexified momenta
\begin{equation}
p_{a\dot{a}}=\lambda_{a}\widetilde{\lambda}_{\dot{a}}%
\end{equation}
for complex spinors $\lambda_{a}$ and $\widetilde{\lambda}_{\dot{a}}$.
We can then speak of a matrix model current $\mathcal{J}_{i}$ for a massless state with a specified
momentum $p_{i}$. The basic dictionary is that a
scattering amplitude is represented as a correlator of currents in the matrix model:
\begin{equation}
i \mathcal{M}_{1,...,n} = \Bigl \langle \mathcal{J}_{1} ... \mathcal{J}_{n} \Bigr \rangle_{\text{MM}}.
\end{equation}
Here the correlator is evaluated by performing the matrix integral while taking the double scaling limit (\ref{scaling}).

This section is organized as follows. First, we begin with a discussion of the flat space limit, and in particular, how to pass
from an abstract correlator of the matrix model to a scattering amplitude. To evaluate such correlation functions, we need to
construct the $\mathbb{CP}^3$ propagator. As we will see, our detailed study of the $\mathbb{CP}^1$ example will give a
good return of investment. Then we construct the asymptotic wave functions.
Finally, as a warmup for our discussion of graviton amplitudes,
we discuss how the model reproduces MHV gluon amplitudes.

\subsection{Flat Space Limit}

\begin{figure}
[ptb]
\begin{center}
\includegraphics[
height=1.8248in,
width=3.902in
]%
{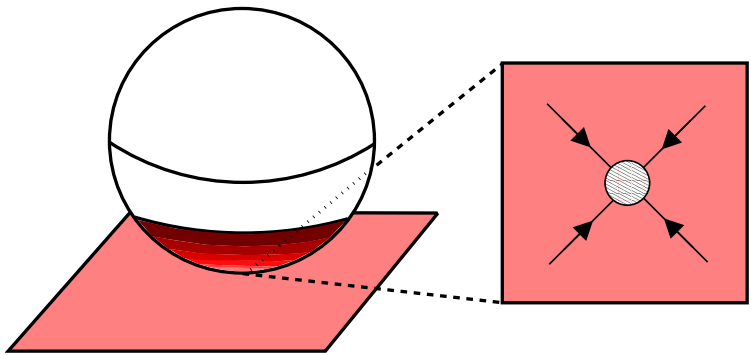}%
\caption{Starting from a round $S^{4}$, the flat space limit is obtained by
zooming in on a small patch near the south pole of the geometry, depicted by
the shaded red region. This is then rescaled, yielding
$\mathbb{R}^{4}$. As depicted in the right panel, scattering amplitudes are
computed by analytically continuing the correlator in the flat space limit to
general complex momenta.}%
\label{flatscatt}%
\end{center}
\end{figure}

In order to compute scattering amplitudes, we need to pass to a 4D theory on flat space-time via the
double scaling limit (\ref{scaling}). In this subsection we discuss in more detail
how to treat this limit. See figure \ref{flatscatt} for a
depiction.

When we zoom in on the region near the south pole region where $x^{\dot a a}$ becomes small, or equivalently,
where $\omega^{\dot a}$ is much smaller than $\pi_a$, the $S^4$ curvature becomes negligible and
the space-time enjoys an effective translation invariance. The hermitian translation
operators are given by the generators $\mathcal{P}$ of $SO(5)$. In the flat space limit, they are related to the generators
$P$ and $K$ via the Wigner-In\"on\"u contraction of the $so(5)$ algebra:
\begin{equation}
\mathcal{P} \rightarrow P + \ellcc^{-2} K
\end{equation}
In this limit, $\mathcal{P}$ remains hermitian, provided the dagger of $P$ is now $P^{\dag} = \ellcc^{-2} K$. In terms of the oscillators, we
rescale the $\omega$ oscillators relative to the $\pi$ oscillators, while keeping the $SO(4)$ subalgebra invariant. This is also
reflected in the equation $\omega^{\dot a} = i x^{\dot a a} \pi_{a} $, so that when $x$ has dimensions of
length, $\omega$ is scaled relative to $\pi$. Since the conformal boost generator leaves
the south pole of the $S^4$ fixed, its effect becomes negligible in the scaling limit (\ref{scaling}).

To make the connection with ordinary translations more explicit, consider the
commutator between the ${\cal P}_{a\dot a}$ generator with the space-time
coordinate operators $X_{\dot a a}$ introduced in Section \ref{sec:spacetime}:
\bea
\left[ {X}_{\dot{a}a},{\cal P}_{\dot{b}b}\right]   \is \epsilon_{ab}\epsilon_{\dot a \dot b}
{X}_0.
\eea
The function space near the south pole region is given by linear combinations of eigenstates of ${X}_{\dot{a}a}/X_0$, as
defined in (\ref{eigenv}), with eigenvalue  $\xX_{\dot a a} \ll 1$. In this region, $X_0$ attains its maximal value, and can be treated like a $c$-
number constant. Hence in the scaling limit, ${\cal P}_{\dot{b}b}$ acts like a translation operator.

Evaluating correlation functions in this limit involves the insertion of the projection
operator $\mathbf{1}_{M}$ into the definition of the amplitudes:
\begin{equation}
\mathbf{1}_{M} \equiv \int d^{4}x\underset{\rho}{%
{\displaystyle\sum}
}\left\vert x,\rho\right) \left( x,\rho\right\vert
\label{projmink}%
\end{equation}
where the domain of integration for $x$ is over the Minkowski patch near the south pole.
Here, the states $\left \vert x , \rho \right)$ are obtained by starting from the
$\mathbb{CP}^{1}$ at the origin, and sweeping out by the $so(5)$ generator $\exp(i x \cdot \mathcal{P})$. Similar considerations hold for
$( x , \rho \vert$. For the most part, such insertions can be ignored. However, when we
turn to a discussion of MHV graviton scattering in section \ref{sec:MHVgrav} where the initial states themselves
disturb the location of the patch (as they are infinitesimal diffeomorphisms), additional care must be taken.

\subsection{$\mathbb{CP}^{3}$ Propagator}

The $\overline{D}$ kinetic operator
(\ref{kinetic}) of the twistor matrix model essentially reduces to the $\mathbb{CP}
^1$ kinetic operator acting
along the twistor lines.
This fact can be anticipated by taking the naive commutative limit of the matrix
model kinetic operator (\ref{contkin}).
This operator acts as a  one-dimensional $\delbar$ derivative along twistor lines.
Hence we expect the continuum limit of the propagator to be delta-function
localized along the directions transverse to this line.

The propagator satisfies the inhomogeneous wave equation with a delta function
source. Following Penrose, we pick this delta function source via the pull back to
the correspondence  space.
Let
\be
\label{newdelta}
\delta(Z,U) = \delta^2(\omega^{\dot a}  - i x^{\dot a a}\pi_a) \,  \delta(\la \pi
\lambda \ra)
\ee
be the delta function that localizes $Z = (\omega^{\dot a},\pi_a )$ at
$U = (x^{\dot a a} \lambda_a, \lambda_a).$
The continuum $\mathbb{CP}^3$ propagator is defined as  the Green's function that solves
\bea
\label{greentt}
\Dbar \Delta(Z,U) \is \delta^3(Z; U)
\eea
It is easily verified that the solution reduces to a projection operator onto the twistor line,
times the $\mathbb{CP}^1$ propagator on this line.
\be
\label{greenttt}
\Delta(Z, U) =  \frac{\delta^2(\omega^{\dot a}\!\!\spc  -\! \spc i x^{\dot a a}\pi_a)}{\la \pi \lambda\ra}
\ee
We now translate this to the non-commutative setting.

The notion of the correspondence space relies on the use of complexified
space-time coordinates, where translations are generated by the operators $P$ rather than
their hermitian counterparts $\mathcal{P}$. In the following, however, we will be interested in the limit in which the $S^4$ gets
very large, that is, we zoom in on a small region near the south pole,  which in the large radius  limit
approaches flat space. In this region, the violation of $SO(5)$ symmetry is minimal. Conversely, the $SO(5)$ generators
act to a very good approximation as translation generators of the Poincare group, which are compatible with
the holomorphic data and do preserve the form of the Green's function (\ref{greenttt}).

Our construction of the Green's function is  modeled after the one employed for
the $\mathbb{CP}^1$ case.
We will use the correspondence
space parametrization and choose an affine coordinate patch $\lambda =
\pi_2/\pi_1$. To every point
labeled by $(x, \lambda)$,
we can associate a coherent state via (\ref{cohersup}) and (\ref{cohertwo}).
Our strategy is to first find the delta function and Green's function that are localized at the twistor line at the
origin $x=0$. We will then find the general solution by acting with the $SO(5)$ symmetry generators.

By analogy with the $\mathbb{CP}^1$ case, we
identify the holomorphic delta function that localizes on
a point $\lambda$ on the twistor line at $x=0$ as $\delta(Z; 0, \lambda) = | 0, \lambda) (0,0|.$
Here $(0,0|$ is the dual north pole state on the ${\mathbb{CP}}^1$ at the origin\footnote{
Here we included a factor of ${1}/{\ell_{pl}^4}$ so that, in the large $N$ limit, the overlap of $(0,0|$ with south pole
states $|0,x)$ at other locations
is properly delta-function normalized:
\be
(0,0| x, 0) = \delta^4(x).
\ee}
\be
(0,0| = {\ell_{pl}^{-4}} \; \la 0 | \pi_1^N.
\ee
Note that both the state $|0,\lambda)$ and $(0, 0|$ are just made up from the $\pi$ oscillators.
So we can view both states as part of the $\mathbb{CP}^1$ Hilbert space at the origin.
We can thus carry over the results of the previous section, and derive that the Green's function,
that satisfies (\ref{greentt}) with the delta function $\delta(Z; 0, \lambda) = | 0, \lambda) (0,0|$,
is given by
$\Delta(Z; 0 ,\lambda) = | 0, \lambda) (0,\lambda|$. Here
$(0,\lambda|$ is the state that satisfies:
\be
(0,\lambda| \epsilon^{ab}\lambda_a \pi_b = (0,0|
\ee
up to terms which are exponentially small at large $N$.
To move away from  the origin, we act by $x \cdot \mathcal{P}$:
\bea
\label{sofive}
|x,\lambda) = e^{ i x \cdot \mathcal{P}}|0,\lambda)  \ \ & ; & \  \
(x,\lambda| = (0,\lambda| e^{- i x \cdot \mathcal{P}}.
\eea
where we assume that the magnitude of $x$ is very small in the flat space limit.

Following the by now familiar pattern, our definition of the delta function (\ref{newdelta}) is
\be
\delta^3(Z; x,\lambda) = |x,\lambda)  (x, 0|
\ee
The right hand side projects onto the position eigenstate $|x,\lambda)$,
as the delta function should; the bra state $(x,0|$ corresponds to the north pole of the $S^{2}$
associated with the affine coordinate system on the twistor line for $x$. In the flat space limit, the state $( x , \lambda |$
satisfies the inhomogeneous holomorphic coherent state condition:
\begin{equation} \label{brann}
(x , \lambda | \epsilon^{ab} \lambda_{a}(\pi_b - i x_{\dot a b} \omega^{\dot a}) = (x , 0|
\end{equation}
up to terms which are exponentially suppressed at large $N$.
How unique is a solution to this equation? We notice that this equation only involves
one linear combination of the $Z^\alpha$ oscillators.
Since there are several oscillators, it might look like this single condition does not
uniquely fix the state.  However, suppose we had found another state that solves eqn (\ref{brann}).
Taking the difference with our solution for $(x,\lambda|$ would yield a bra state $(\psi|$ that is
annihilated by a linear  combination of annihilation operators.
Clearly no such state exists. Hence our solution is unique up to small correction terms.

The propagator that satisfies (\ref{greentt}), with the above identification of the
delta function, is now immediately found to be
\be
\label{cpprop}
\Delta 
(Z; x,\lambda) = |x,\lambda)(x,\lambda|\, .
\ee
Verification of the Green's function property follows
immediately from the fact that $[\overline{D} , \mathcal{P}] = 0$. Indeed,
since $\overline{D} ( |0 , \lambda) (0 , \lambda | ) = |0 , \lambda) (0 , 0|$,
we obtain:
\begin{equation}
\label{cpverif}
\overline{D} ( |x,\lambda ) (x, \lambda| ) = \overline{D} (e^{i x \cdot \mathcal{P}} |0,\lambda ) (0,\lambda| e^{- i x \cdot \mathcal{P}} ) = |x,\lambda ) (x, 0|
\end{equation}
The final expression (\ref{cpprop}) for the $\mathbb{CP}^3$ propagator will
be used repeatedly in the following sections for the computation of
scattering amplitudes. Because $\overline{D}$ is an invertible
map on this basis of matrices, we can also invert both this map and the action by $\exp(i x \cdot \mathcal{P})$.
This establishes the uniqueness of the Green's function solution.

A last piece of information we need is the inner product between the special bra
states $(x,\lambda|$ with a position eigenstates. Using the earlier calculation of the overlap of position eigenstates,
we find that
\bea
\label{overlaprop}
( x_1, \lambda_1 \spc |\spc x_2, \lambda_2) \is \, \delta^4\bigl(x_{12}\bigr) \, ( \lambda_1 | \lambda_2\spc )
\, = \, \frac{ \delta^4\bigl(x_{12}\bigr) } {\la
\spc \lambda_1\spc \lambda_2 \ra}
\eea
which should be compared with the continuum version (\ref{greenttt}) of the $\mathbb{CP}^3$ Green's function.

The generalization to supertwistor space is straightforward. Starting from the $\mathbb{CP}^{1}$ at $x = \theta = 0$, we have the delta function
on the $\mathbb{CP}^{1}$, $| 0 , 0 , \lambda )( 0 , 0 , 0|$. At small $x$ and $\theta$, this corresponds to a point
on the supercorrespondence space ${\cal U} = \bigl(x^{\dot a a}\lambda_a, \theta^{ia} \lambda_a, \lambda_a\bigr)$. In the flat space limit, the
state $(x , \theta , \lambda|$ satisfies:
\bea
(x,\theta, \lambda| \epsilon^{ab}\lambda_a(\pi_b - i x_{\dot a b} \omega^{\dot a} - \eta_{ij} \psi^i \theta^j_b) = (x,\theta,
0|
\eea
Acting by a symmetry generator of $S^{4|8}$, we can move out to a general value of
$x$ and $\theta$. This symmetry generator commutes with the supersymmetric kinetic operator $\overline{\mathcal{D}}$,
so again the verification of the Green's function property is trivial.
The propagator is then given by
\be
(x_1,\theta_1,\lambda_1|x_2,\theta_2, \lambda_2 ) = \frac{\delta^{4|8} (x_{12})}{\la \lambda_1\lambda_2\ra}
\ee
where the delta function is over the $\mathcal{N} = 4$ superspace.

\subsection{Space-Time Currents}

The next step in assembling the ingredients of the S-matrix is to determine a physically natural basis of
currents. These are specified by a choice of background gauge
field $\mathcal{A}^{\alpha} = Z^{\alpha} V$ for some $V = V_\aA \otimes \tau^\aA$ and $\tau^{\aA}$
a generator of $u(N_c) \times gl(k_N)$. The generators $u(N_c)$ define currents on color space,
while the $gl(k_N)$ generators are deformations of the geometry itself. At a heuristic level, we are
interested in taking $V = V(Z)$ to be a ``locally holomorphic'' function of just the $Z$'s
in the sense that they commute with the holomorphic coordinates. In this sense, such functions do
not disturb the holomorphic geometry of twistor space.

Strictly speaking, this cannot really be done on a finite size $S^{4}$, and in particular in the
finite $N$ theory. The reason is that all matrices we write
down will be a power series in both $Z$ and $Z^{\dag}$, so all currents will inevitably distort the geometry. Indeed,
it is precisely this feature which suggests a connection with gravity. This is closely related to the presentation of the
position eigenstates $\vert x , \lambda)$ in the flat space limit. Recall that in this limit, $\vert x , \lambda)$
is obtained by starting from the south pole state $\vert 0 , \lambda)$ and applying
$\exp(i x \cdot \mathcal{P})$. This intrinsically links this collection of states to a small neighborhood in the vicinity of the south pole.
On the finite size $S^{4}$, we could have alternatively started from the north pole and rotated by a different $SO(5)$ rotation to reach the same point on the
$S^{4}$. Note, however, this operation would not have been holomorphic in the original $x^{\dot a a}$,
as it involves transport from the point at infinity. Hence, when we work at finite $N$, the most
we can hope for is an approximate notion of holomorphy in the $V(Z)$ which becomes exact in the large
$N$ limit. When we turn to a discussion of plane wave solutions, we shall give a more precise characterization
of such ``locally holomorphic'' $V$'s.

In the following, we  will distinguish two special classes of $V$ generators.
The first class are the closest analogue of local color gauge rotations acting on $\tQ$.
\bea
V^\aA (Z) \is \tau^\aA\, V(Z)
\eea
We can call these transformations local color rotations, because they act on the
color index of $\tQ$ but otherwise commute with the holomorphic coordinates (at least locally).
They therefore do not induce any coordinate shift of the holomorphic coordinates. As we will see, the correlation
functions of currents associated with  this class of transformations will correspond to gauge theory MHV amplitudes.

A second special class of transformations are those that leave the color index unchanged, but act non-trivially on
the holomorphic coordinates $Z^\alpha$ by means of an infinitesimal $gl(k_N)$ transformation.
A natural class of generators are \cite{TMMII}:
\bea\label{VAAZ}
V_{\dot a a}(Z) \is  \, \mathcal{P}_{\dot a a} V(Z)
\eea
where $\mathcal{P}_{\dot a a} = P_{\dot a a} + K^{\dot a a}$ is an $so(5)$ generator.
The lefthand side $V_{\dot a a}(Z)$ is a $gl(k_N)$ generator which
contains a single $Z^\dag$ oscillator. Via the commutator, it
describes a holomorphic vector field on the non-commutative
$\mathbb{CP}^3$. Correlation functions of currents associated with this
class of transformations will correspond to gravity MHV amplitudes.

In the supersymmetric case there are additional transformations and associated currents.
These are given by the purely fermionic $su(4)$ generators, as well
as mixed bosonic and fermionic currents. The former can be identified with the gauged
R-symmetry of a supergravity theory with $\mathcal{N} = 4$ supersymmetry,
while the fermionic components correspond to the gravitinos.

\subsubsection{Plane Waves}

To complete our characterization of the $V$'s, we now construct
operators corresponding to asymptotic states with specified complexified
momentum $p_{a \dot a} = \lambda_{a} \tilde{\lambda}_{\dot a}$, as appropriate for a discussion of
scattering theory in twistor space.

The construction of the solutions in the flat space limit is obtained by viewing $\vert x , \lambda ) (x , \lambda |$ as a designated projection to a point of the correspondence space. This is of course in accord with the identification of the flat space limit projection matrix $\mathbf{1}_{M}$ of equation (\ref{projmink}). A momentum eigenstate is then given by summing over the continuum position $x$ of the operator $e^{i p \cdot x} \vert x , \lambda ) (x , \lambda |$ in this small patch.

We now formalize the algebraic conditions for a matrix $V$ to be a
momentum eigenstate. The first requirement is that the propagating mode $V$ is specified by
an asymptotic source $f$ via $I_{\alpha \beta} \mathcal{A}^{\alpha} Z^{\beta} = f$, for $f$ a $k_{N} \times k_{N+2}$ matrix.
Dropping all group theory indices, the identification $\mathcal{A}^{\alpha} = Z^{\alpha} V$ is summarized by the condition:
\begin{equation}\label{equSOURCE}
\overline{D} V = f.
\end{equation}
Here, $f$ is treated as an a priori arbitrary source for the wave function $V$.

To construct a momentum eigenstate, we now further restrict attention to
operators $V(p)$ with a specified momentum $p_{\dot a a}$.
In the flat space limit, this is designated by the condition:
\begin{equation}\label{eigenMOM}
[P_{\dot a a} , V(p) ] = p_{\dot a a} V(p)
\end{equation}
where $V(p)$ is viewed as a state in the large $N$ limit of the adjoint representation of $gl(k_{N+1})$.\footnote{ At finite $N$, this condition has various correction terms. Indeed, whereas the $SO(5)$ generator $\mathcal{P}$ is Hermitian, $P$ is nilpotent. In the flat space limit, however, this is not much of an issue.}
Note that since $\overline{D}$ and $\mathcal{P}$ commute, we can simultaneously impose equations (\ref{equSOURCE}) and (\ref{eigenMOM}). In this case, we write $f(p)$ for a source of momentum $p$.

We now construct the form of these solutions in the flat space limit. To do this, we briefly review the construction of solutions to the free wave equation in commutative twistor space. Solutions to the helicity $h$ free field wave equation on 4D spacetime are conveniently specified by the Penrose transform. The basic idea is to look for elements of $H^{1}(\mathbb{PT}^{\prime} , \mathcal{O}(2 h - 2))$, where  $\mathbb{PT}^{\prime} = \mathbb{CP}^{3} - \mathbb{CP}^{1}_{\infty}$ is projective twistor space with the line at infinity deleted. Given a cohomology representative $f_p(\omega , \pi)$ on $\mathbb{PT}^{\prime}$, we can via the twistor equation $\omega = i x \pi$
obtain a representative on correspondence space $\mathbb{C}^{4} \times \mathbb{CP}^{1}$, with coordinates $(x^{\dot a a},s_{a})$.
Observe that this is not an arbitrary section on the correspondence space; It satisfies the condition:
\begin{equation}\label{wavequ}
\lambda^{a} \partial_{\dot a a}f_{p}(x , s) = 0
\end{equation}
The Penrose transform amounts to a contour integral over
the $\mathbb{CP}^{1}$ factor, resulting in a 4D space-time field:
\begin{equation}\label{phidef}
\phi(x) = \oint \langle s d s \rangle f_{p}(x , s)
\end{equation}
where here, we have specialized to the case where $f_{p}$ is a degree zero $(0,1)$-form.
Other helicities are covered by including factors of $\pi$ or $\partial / \partial \omega$ acting on the integrand $f_p$. The
plane wave solution is:
\begin{equation}
f_{p}(x , s) = \exp(i p \cdot x) \delta(\langle s \lambda \rangle)
\end{equation}
where the delta function is a $(0,1)$ form of specified homogeneity.
The Fourier transform of equation (\ref{phidef}) then provides a helicity $h$
momentum eigenstate.

Let us now turn to the fuzzy setting. The analogue of the delta function $\delta(\langle s \lambda \rangle)$ on a fuzzy $\mathbb{CP}^{1}$ is the operator
$| \lambda ) ( 0 |$. Summing over a basis of position states, much as in our discussion of the projection to Minkowski space,
the corresponding operator $f(p)$ is:
\begin{equation}\label{fpdef}
f(p) = \int d^{4} x  \exp(i p \cdot x) | x , \lambda ) (x , 0|
\end{equation}
where $p_{\dot a} = \lambda_{a} \widetilde{\lambda}_{\dot a}$.
Observe that this operator formally satisfies the operator equation $\pi^{a} [P_{\dot a a} , f(p)] = 0$,
which is the analogue of equation (\ref{wavequ}). To obtain the corresponding plane
wave operator $V(p)$, we need to integrate equation (\ref{equSOURCE}). Here
we can make use of the ${\mathbb{CP}^3}$  propagator (\ref{cpprop}), which satisfies (\ref{cpverif}). We thus arrive at the following
definition of the plane wave operators:
\begin{equation}\label{Vop}
V(p) = \int d^{4} x \,\, e^{i p \cdot x} |x , \lambda ) ( x , \lambda |.
\end{equation}
The integral expressions  (\ref{fpdef}) and (\ref{Vop}) should both be viewed within the context of the scaling limit
(\ref{scaling}). This means that the integral runs over a local patch near $x=0$,
while the momentum $p$ is scaled accordingly (see figure \ref{flatscatt}).

The generalization to the supersymmetric case is straightforward. The plane wave states then also
depend on anti-commuting variables $\zeta$
\bea
\label{vmoms}
V({p,\zeta}) =  \int d^{4|8} x \, e^{ip \cdot x+ \zeta \cdot \la\lambda \theta\ra}\spc  | x,\theta, \lambda) (x, \theta,\lambda|
\eea
The components of the $\mathcal{N} = 4$ supermultiplet $V$ provide states of different helicity in the 4D theory.
For the $\mathcal{N} = 4$ vector multiplet, the bottom component constitutes a plus helicity gluon and
the top component is a minus helicity gluon. For $\mathcal{N} = 4$ gravitons, the plus
helicity and minus helicity gravitons sit in two different supermultiplets, with the plus
helicity mode at the bottom of its multiplet, and the minus helicity mode at the top of its multiplet.
As noted in \cite{Nair:1988bq}, MHV scattering amplitudes in $\mathcal{N} = 4$ superspace lead to a kinematic
factor of $\langle n 1 \rangle^{4}$ for minus helicity states of momenta $p_1 = \tilde{\lambda}_{1} \lambda_{1}$
and $p_n = \tilde{\lambda}_{n} \lambda_{n}$. To avoid clutter, we shall often leave implicit the integration over $\mathcal{N} = 4$ superspace.

\subsection{MHV\ Gluon Scattering}

We are now ready to compute amplitudes. We first consider the MHV\ gluon scattering amplitudes \cite{Parke:1986gb}.
The computation is similar to that in \cite{Nair:1988bq} and to the twistor string theory calculation in \cite
{Witten:2003nn}. As discussed in section 3.2, the gaussian matrix model comes with a natural set of currents ${\cal J}(V)$, given in
(\ref{jvee}), that describe the coupling to a gauge field $\bA$, defined on non-commutative $\mathbb{CP}^3$. Here the $V$ are
linear maps acting on the $\mathbb{CP}^3$ Hilbert space, and should be thought of as the asymptotic wave
functions of the gluon states. A supermultiplet of a gluon state with color charge $\tau^{\aA}$ and supermomentum $(p,\zeta)$ is described
by
\bea
\label{jgauge}
{\cal J}^{\aA}(p , \zeta) = \spc  \spc {\rm Tr} \Bigl(\tau^{\aA} V(p , \zeta)\spc {\qQ}\spc \Dbar{\tQ} \Bigr)
\eea
where $V(p,\zeta)$ is the momentum eigenstate defined in (\ref{vmoms}).

The scattering amplitude is now directly given by the matrix model expectation value
\be
i {\cal M}_{1,...,n}= \Bigl\langle {\cal J}_{1}(p_1,\zeta_1)\spc ...\, {\cal J}_{n}(p_n,\zeta_n)\Bigr\rangle_{\rm
MM}.\label{ampdef}%
\ee
As explained in section 3.2, performing the matrix integral is trivial: the vertex operators (\ref{jgauge}) represent the
response to a simple field redefinition $\delta_V{Q} = V Q, \delta_V \tQ = 0$. Performing resulting Wick contractions
immediately leads to the following expression for the color-stripped subamplitude
\be
A_{1,...,n}= {\rm Tr}\Bigl(V({p_1,\zeta_1}) \ldots V({p_n,\zeta_n}) \Bigr)
\ee
Here the trace is over the Hilbert space ${\mathbb{CP}^{3|4}}$ of the non-commutative supersymmetric Hilbert
space.

The rest of the calculation is equally straightforward. Inserting the definition (\ref{vmoms}) for the plane wave
operators, we first evaluate the amplitude in position space
\bea
\widetilde{A}_{1,...,n} \is\underset{i=1}{\overset{n}{%
{\displaystyle\prod}
}}\bigl(\, x_i\spc ,  \theta_i, \lambda_i \, \bigr| \spc  x_{i + 1}, \theta_{i+1}, \lambda_{i+1} \bigr)
\, = \, \underset{i=1}{\overset{n}{%
{\displaystyle\prod}
}}\frac{\delta^{4|8}(x_{i-1,i})}{\left\langle \lambda_{i-1}\spc \lambda_{i}\right\rangle }%
\eea
where $i=0$ is identified with $i=n$.  Here we used the result (\ref{overlaprop}) for the overlap. %
Fourier transforming back to momentum space, we obtain
\bea
{\cal A}_{1,...,n}  \is \delta^{4}\Bigl(  \underset
{i=1}{\overset{n}{%
{\displaystyle\sum}
}} 
p_i\Bigr)
\frac{\left\langle n1\right\rangle ^{4}}{\left\langle 12\right\rangle
...\left\langle n1\right\rangle }%
\eea
which we recognize as the color-stripped contribution to the gluon MHV amplitude. Let us note that the form of this amplitude is fixed up to an overall multiplicative constant. In particular, there is an overall finite factor of $\delta(0)$, which is connected with a short distance cutoff. Here, this simply specifies a reference energy scale for the external momenta $p_i$ of the amplitude.

\section{MHV\ Graviton Scattering}

\label{sec:MHVgrav}

In this section we show how MHV graviton amplitudes are computed in the matrix model.
MHV graviton scattering amplitudes involves a pair of minus helicity gravitons
(labeled by 1 and $n$) and an arbitrary number of ingoing plus helicity
gravitons (numbered $2$ to $n-1$). In this section we demonstrate that this
amplitude is given by the expectation value of gravitational currents in the matrix model:
\begin{equation}
\label{gravcorrelator}i\mathcal{M}_{\mathrm{MHV}}=\Bigl\langle \widetilde{\mathcal{T}}%
_{1}(p_{1},\zeta_{1})\hspace{1pt}\mathcal{T}_{+}(p_{2}
)\cdots\mathcal{T}_{+}(p_{n-1})\hspace{1pt}\hspace{1pt}
\widetilde{\mathcal{T}}_{n}(p_{n},\zeta_{n}) \Bigr\rangle_{\text{MM}}
\end{equation}
The twistor realization of this calculation builds on Penrose's non-linear
graviton construction of (anti-)selfdual space time backgrounds. Collectively,
the plus helicity gravitons can be thought of as representing a selfdual
background geometry, on which the minus helicity graviton propagates. Upon
reversing the momentum of one of the minus helicity gravitons (say, the one
labeled by $n$), the MHV scattering amplitude represents \cite{Mason:2008jy}
the process of an incoming minus helicity graviton bouncing off the selfdual
background geometry, while undergoing an helicity-flip. The MHV amplitudes
arise by expanding out the background field in terms of linearized
perturbations around flat space-time.

\begin{figure}[ptb]
\begin{center}
\includegraphics[
height=2.4388in,
width=6.4472in
]{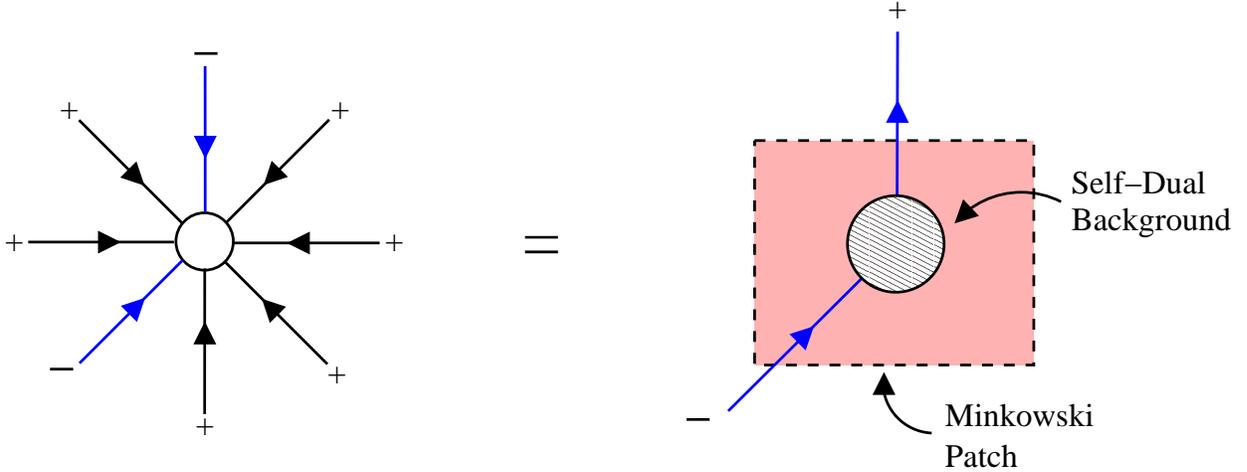}
\end{center}
\caption{The MHV graviton scattering amplitude represents the process of a
minus helicity graviton reflecting off a self-dual background, built up by
the metric fluctuations with plus helicity. The scattering produces a
helicity flip. In the matrix model this amplitude is computed in a small ``Minkowski patch''.
Figure modified from \cite{Mason:2008jy}.}%
\label{sdscatt}%
\end{figure}

Graviton excitations are associated with deformations
of space-time. We have already identified a natural class of $gl(k_N)$ generators in section 3.2
given by equation (\ref{VAAZ}):
\begin{equation}\label{TVVV}
T_{\vvv}(p , \zeta)  = (v \cdot \mathcal{P})\hspace{1pt} V(p,\zeta)
\end{equation}
in the obvious change of notation. Here $v\cdot\mathcal{P}=v^{\dot{a}a}P_{\dot{a}a}+v_{\dot{a}a}K^{\dot{a}a}$ is
the (complexified) $SO(5)$ generator, that in the local Minkowski patch around
$\omega=0$ acts like a translation generator in the direction $v^{\dot{a}a}$. These are
momentum eigenstates connected with translations.
Note, however, that it also contains a component proportional to the conformal
boost generator.

Before explaining the geometric content of these currents, let us first
specify the form of the tensor $v^{\dot a a}$ appearing
in $T_{\vvv}$. Here we will slightly simplify our task. Since we are anticipating an interpretation in
terms of graviton states, we can carry over the standard space-time treatment
of the associated polarization tensors. Of the four possible polarizations
$v^{\dot a a}$, we can eliminate two due to spin 2 gauge invariance:
linearized space-time diffeomorphisms shift $v^{\dot a a}$ with an amount
proportional to the momentum $p^{\dot a a}$. Such longitudinal modes decouple
from the amplitude (we will make this explicit shortly). This then allows us
to choose the transversality condition $v_{\dot a a}p^{\dot a a} = 0$. The
remaining two polarizations split up into a positive and minus helicity
component via
\begin{equation}
\label{vplus}v^{+}_{\dot a a} = \tilde\lambda_{\dot a}\mu_{a}, \qquad
\qquad\langle\mu\lambda\rangle= 1
\end{equation}
\begin{equation}
\label{vminus}v^{-}_{\dot a a} = \tilde\mu_{\dot a}\lambda_{a}, \qquad
\qquad[\hspace{1pt} \tilde\mu\hspace{1pt} \tilde\lambda\hspace{1pt} ] = 1
\end{equation}
This choice of the dual two component spinors $\mu_{a}$ and $\tilde\mu_{\dot a}$ is
not unique, due to the residual on-shell gauge invariance $\mu_{a} \to\mu_{a}
+ \lambda_{a}$ and $\tilde\mu_{\dot a} \to\tilde\mu_{\dot a} + \tilde
\lambda_{\dot a}$. We will make a specific choice for $\mu_{a}$ and $\mu_{\dot
a}$ later on. Note that the transversality condition $p^{\dot a a} v_{\dot a
a} = 0$ has the natural consequence that, in the flat space limit, the
generator $v^{\dot a a} \mathcal{P}_{\dot a a} \rightarrow v^{\dot a a}
P_{\dot a a}$ commutes with $V(p,\zeta)$. To denote the two types of helicities, we will
continue to write $T_{+}(p)$ or just $T(p)$ for the plus helicity gravitons, while we will write
$T_{-}(p,\zeta)$ or just $\tilde{T}(p,\zeta)$ for the minus helicity gravitons.

It is instructive to look at the leading order form of the plus helicity graviton currents
in the local Minkowski region near $\omega=0$. In this flat space limit, the
polarization operator of the plus helicity gravitons takes the form
\begin{equation}
(v_{k}^{+})^{\dot{a}a}P_{\dot{a}a}=\langle\mu_{k}\pi\rangle\lbrack
\tilde{\lambda}_{k}\hspace{1pt}\omega^{\dag}]=[\tilde{\lambda}\hspace
{1pt}\omega^{\dag}] \label{poshift}%
\end{equation}
where we used that $\pi^{a}V(p_{k})=\lambda_{k}^{a}V(p_{k})$ and $\langle
\mu_{k}\hspace{1pt}\lambda_{k}\rangle=1$. Here the invariance under shifts
$\mu_{a}\rightarrow\mu_{a}+\lambda_{a}$ is made manifest. Due to the
non-commutativity, equation (\ref{poshift}) represents a holomorphic
derivative along the $\omega_{a}$ directions. The vertex operators of the
plus helicity gravitons can thus be viewed as living in the holomorphic
tangent space to twistor space. Moreover, we learn that the insertion of the
plus helicity generator $T_{+}(p)$ represent small transverse deformations of the local
twistor line, via an infinitesimal diffeomorphism that shifts $\omega_{\dot
{a}}$ by an amount proportional to $\tilde{\lambda}_{\dot{a}}V(p)$.
Similar considerations hold for the minus helicity generator $T_{-}(p,\zeta)$.

Having specified a class of $gl(k_N)$ elements, we next specify the corresponding
vertex operators. As we have already mentioned, the $gl(k_N) \times \widetilde{gl}(k_N)$ gauge symmetry are associated
with the bulk gauge field $\mathcal{A}$ and the compensator $\widetilde{\mathcal{A}}$. As opposed to the case of MHV gluon scattering,
in MHV graviton scattering the two helicity modes fill out distinct $\mathcal{N} = 4$ supermultiplets.
This means that we should anticipate that the presentation of their vertex operators can be different.
In an MHV amplitude, we expect the plus helicity modes to deform the local geometry,
while the two minus helicity modes specify asymptotic data. This motivates our specification of the vertex operators:
\begin{equation}\label{gravv}
\mathcal{T}_{+}(p) =\mathrm{Tr}\Bigl(  \bigl[T_{+}(p) , {Q} \bigr] \overline{D}{\tQ}   \Bigr)
\end{equation}
Here, the plus helicity currents $\mathcal{T}_{+}$ act via the algebra of vector fields, and embody the result of performing an
adjoint gauge variation
\bea
\delta Q = \bigl[ T_{+}(p), Q\bigr] \ & ; & \ \delta\widetilde{Q}=0\label{deform}%
\eea
Geometrically, these transformations build up a selfdual background off of which a minus helicity graviton can scatter.
The minus helicity gravitons specify flow into and out of the patch near the south pole, and transform in a dual representation
on which the vector fields can act.

This identification makes direct contact with the continuum twistor description of graviton
scattering amplitudes \cite{Mason:2008jy}. In Penrose's non-linear graviton
construction \cite{Penrose:1976jq}, a selfdual background space-time is converted into a twistor
space with a deformed complex structure. The deformed space is still described
by the same coordinates as usual twistor space, and $\pi_{a}$ still represents
a holomorphic coordinate along the twistor lines. The coordinates
$\omega^{\dot a}$ are however no longer holomorphic in the distorted complex structure.

To bring out this geometric picture, let us introduce the self-dual gravity
background given by the formal exponentiation of the plus helicity metric
fluctuations%
\begin{equation}
\exp\left(  T_{+}(h) \right)  =\prod_{k=2}^{n-1}\hspace{1pt}\exp\left(
h_{k}T_{+}(p_{k} ) \right)
\end{equation}
where $h_{k}$ is the infinitesimal amplitude of the $k$-th mode. Each
individual graviton amounts to a small chiral $GL(k_{N})$ gauge transformation, which acts via commutation as in
(\ref{deform}). This transformation represents the non-linear graviton
background. We can formally write the action of the gaussian model propagating
in a self-dual space-time background as%
\begin{equation}
S_{h}(Q,\tilde{Q})=\mathrm{Tr}\left(  e^{T_{+}(h)} \hspace{1pt} Q\overline{D}\widetilde
{Q}\right)
\end{equation}
where the expansion of the exponential acts by all possible orderings of nested commutators on $Q$.
As explained, the MHV graviton scattering amplitude can be obtained by
considering the propagation of a minus helicity graviton on top of this
self-dual background.

Consider next the vertex operators for the minus helicity gravitons. As opposed to the case of MHV gluon scattering,
in MHV graviton scattering the two helicity modes fill out distinct $\mathcal{N} = 4$ supermultiplets. This means that we should anticipate that the presentation of their vertex operators may be different.
As we have seen, the plus helicity modes are identified with holomorphic vector fields that deform the local geometry. Hence, following the continuum twistor description,
it is natural to view the minus helicity modes as being in the cotangent space, which
are transformed by this algebra of vector fields. We will identify the minus helicity graviton with the following currents
\begin{equation}
\label{neghel}
\qquad \widetilde{\mathcal{T}}_{-}(p_s , \zeta_s) = \mathrm{Tr}\Bigl(  \tilde{T}_{-}(p_{s} , \zeta_{s})  {Q} \overline{D}{\tQ}   \Bigr), \qquad s=1,n
\end{equation}
with $\tilde{T}_{-}$ as in equation (\ref{TVVV}), with polarization tensor appropriate for a minus helicity graviton.
The momentum associated with the minus helicity gravitons flows in and out of the local Minkowski patch, as indicated in fig. \ref{sdscatt}.

The geometric interpretation of (\ref{neghel}) as associated with cotangent factors arises as follows. The polarization factor of the
minus helicity graviton, say graviton 1, contains the contraction of
$(v_{1}^{-})_{\dot a a}$ with the conformal boost generator $K^{a\dot a}$,
\begin{equation}
\label{veen}(v_{1}^{-})_{\dot a a} K^{\dot a a} = \langle\lambda_{1} \pi
^{\dag}\rangle[\tilde{\mu}_{1} \hspace{1pt} \omega].
\end{equation}
When we compute the amplitude, we will see that this operator naturally
pairs with the holomorphic tangent vectors (\ref{poshift}), in the precise way
that one would expect from an element of the cotangent space.

\subsection{Evaluation of the Graviton Amplitude}

We will now calculate the matrix model expectation value (\ref{gravcorrelator}%
) and show that it reproduces the MHV amplitude (\ref{BGK}). Our notation
for the plus and minus helicity currents is
$\mathcal{P}_{k}=(v_{k}^{+}\!\cdot\!\hspace{1pt}\mathcal{P})$ , for
$k=2,..,n-1$, $\tilde{\mathcal{P}}_{s}=(v_{s}^{-}\!\cdot\!\hspace
{1pt}\mathcal{P})$ for $s=1,n$, and $V_{\ell}=V(p_{\ell})$ for all
$\ell=1,..,n$.

\begin{figure}[ptb]
\begin{center}
\includegraphics[
height=1.5in,
width=2.2in
]{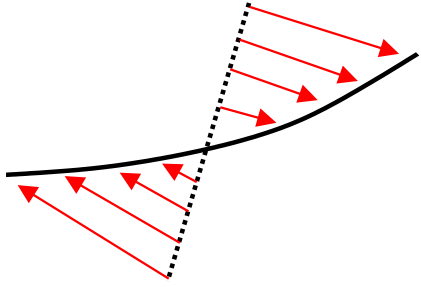}
\end{center}
\caption{Depiction of the non-linear graviton (red arrows) in fuzzy twistor
space. The dashed line indicates a $\mathbb{CP}^{1}$ with respect to a fixed
complex structure. The curved line indicates the deformation due to the
non-linear graviton. Each red arrow corresponds to a \textquotedblleft
bit\textquotedblright\ of the background and the cumulative effect generates a
finite complex structure deformation. See \cite{Mason:2008jy} for a related
depiction of the non-linear graviton.}%
\label{nlingrav}%
\end{figure}

The calculation is remarkably straightforward. The basic form of
the combinatorics is similar to the computation in
\cite{Mason:2008jy} and also has some overlap with the proposal of
\cite{Nair:2005iv}. Using the exponentiated form of the plus helicity gravitons as
representing a self-dual background, performing the Wick contractions produces
the following amplitude%
\begin{equation}
\mathrm{Tr}\Bigl(\tilde{T}_{1} \cdot\exp\bigl(T_{+}%
(h) \bigr)\cdot\tilde{T}_{n}\Bigr)
\end{equation}
This expression exhibits the geometric interpretation of the amplitude as the
propagation of graviton $1$ through the self-dual background, emerging as an
outgoing graviton $n$. See figure \ref{nlingrav} for a depiction of how the
non-linear graviton deforms the local twistor geometry.

One could in principle try to do the calculation with the fully exponentiated
background. Instead, let us expand the self-dual background into the
individual graviton modes. Expanding out each contribution from $T_{+}$, we obtain a
set of nested commutators for our subamplitude:
\begin{equation}
A_{1,...,n}= \mathrm{Tr}\left( \bigl[ [ \tilde{T}_{1} , T_{2} ] ,... , T_{n-1} \big] \cdot \tilde{T}_{n} \right)  .\label{subampgrav}%
\end{equation}
The continuum theory  interpretation of this form of the amplitude is hopefully clear; The nested set of commutators correspond to the action of Lie derivatives which act to the left on $\tilde{T}_{1}$. Later, we will see that $\tilde{T}_{1}$ can be viewed as transforming in the cotangent bundle. The successive Lie derivatives then
return a modified cotangent vector which dots with the (dualized) cotangent vector $\tilde{T}_{n}$. The
full amplitude is then given by summing over $P_{(2,...,n-1)}$, that is, all permutations of the plus
helicity gravitons.\footnote{Returning to equation (\ref{gravcorrelator}), we should note here that Wick's
theorem also produces additional orderings of the currents where
$\tilde{T}_{1}$ and $\tilde{T}_{n}$ are not
adjacent to one another. These do not correspond to graviton scattering, but
rather describe a graviton which would be absorbed into the self-dual
background. In the flat space limit, all of these contributions vanish
simply due to conservation of helicity and momentum.}
 Finally, in the above expression, we have suppressed an overall prefactor of
$\kappa = \sqrt{16 \pi G_N}$ for each graviton vertex operator. The precise value of
Newton's constant cannot, however, be fixed by just MHV graviton scattering.
Instead it requires a discussion of how the gaussian matrix model embeds inside a
more complete framework of the type discussed in the companion paper \cite{TMMII}, that include modes that propagate between different twistor lines.

We would like to evaluate the subamplitude $A_{1,...,n}$, while taking the
scaling limit (\ref{scaling}). As opposed to the case of gluon scattering,
implementing this scaling limit for graviton scattering is slightly more
subtle, because the currents are themselves related to infinitesimal
translations. Indeed, the correlator we are considering is technically defined
with respect to an $S^{4}$ geometry, while the scattering amplitude is defined
by taking a scaling limit in a small patch near the south pole. Roughly
speaking, we need to take into account the fact that whereas in Minkowski
space we can have a momentum flow in from infinity, on the compact $S^{4}$
geometry, this requires introducing a source and a sink.

A convenient way to capture the large radius behavior of the correlator is as
follows. The flat space limit is obtained via insertions of the operator
$\mathbf{1}_{M}$ introduced in equation (\ref{projmink}). For the most part, these insertions can be ignored.
However, the insertions involving the minus helicity states must be treated with more care, because they can move
the domain of definition which we are zooming in on. Focussing on the insertion
between the two minus helicity graviton currents, we can insert a factor of $\mathbf{1}_{M}$ into the subamplitude:
\begin{equation}
A_{1,...,n}= \mathrm{Tr}\left( \bigl[ [\tilde{T}_{1} , T_{2} ] ,... , T_{n-1} \bigr] \cdot \tilde{T}_{n} \spc \mathbf{1}_{M}\right)  .\label{subampgrav}%
\end{equation}
which implements the flat space limit. Recall from equation (\ref{projmink}), the projection to the
patch is obtained by a sum over $\vert x , \rho \rangle \langle x , \rho \vert$. Here we adopt angular
brackets to indicate we are now working about a patch of Minkowski space. To leading order
in the flat space limit, the translation generators annihilate this state via right multiplication:
$\vert x , \rho\rangle \langle x , \rho \vert P_{\dot a a} = 0$.

We now consider the subamplitude $A_{1,...,n}$ in this scaling limit. To simplify the expression,
we first expand out each successive commutator. Doing so, we see that
for the plus helicity generators $T_{i}$, $\vert x , \rho\rangle \langle x , \rho \vert T_{i} = 0$. In other words,
each nested commutator can be replaced by ordinary matrix multiplication. To
represent the flow of the momentum into the patch, we keep $\tilde
{\mathcal{P}}_{1}$ to the left of $V_{1}$ and we take $\tilde{\mathcal{P}}%
_{n}$ to the right of $V_{n}$, which can be done since $[\tilde{\mathcal{P}%
}_{n},V_{n}]\rightarrow0$ in the flat space limit. Our new task is therefore
to evaluate:
\begin{equation}
\left\langle
x,\rho\right\vert \tilde{\mathcal{P}}_{1}V_{1}\cdot P_{2}V_{2}\cdots
P_{n-1}V_{n-1}V_{n}\tilde{\mathcal{P}}_{n}\left\vert x,\rho\right\rangle
\end{equation}
where all operators now compose via ordinary matrix multiplication. In
this expression we have made the substitution $T_{i} \rightarrow P_{i} V_{i}$
for all the plus helicity generators, as appropriate in the flat space limit.
This form of the subamplitude has the clear interpretation of the
minus helicity graviton flowing into the local patch near the south
pole, scattering off a self-dual background, and then flowing out of the local Minkowski patch.

The basic plan of the calculation is therefore very simple: we first transport
the $k=2,...,n-2$ plus helicity factors of $P_{k}$ all the way to the left.
Along the way we collect all contributions of the commutators with the plane
wave operators $V(p_{l})$%
\begin{equation}\label{leadingpart}
\left[  P_{k}, V(p_{l})\right]  =(v_{k}^{+}\!\cdot\!\hspace{1pt}p_{l}%
)\hspace{1pt}V(p_{l})\,.
\end{equation}
Note that in the limit we are considering, these momentum factors give the dominant contribution:
the contributions from $\langle x , \rho \vert P_{k}$ and from moving $P_{k}$ past $\tilde{\mathcal{P}}_{1}$ are
both negligible.
So the leading order result is:%
\begin{equation}
\left\langle
x,\rho\right\vert \tilde{\mathcal{P}}_{1}V_{1}V_{2}\hspace{1pt}...V_{n-2}%
P_{n-1}V_{n-1}V_{n}\tilde{\mathcal{P}}_{n}\hspace{1pt}\left\vert
x,\rho\right\rangle \text{ }\prod_{k=2}^{n-2}v_{k}^{+}\ccdot\left(
p_{1}+...+p_{k-1}\right)  .\label{intermedone}%
\end{equation}
Note that alternatively, we could have allowed all of the $P$'s to pass to the right.
This would result in factors of $v_{k}^{+}\ccdot\left(  p_{k+1}+...+p_{n}%
\right)  $. Via conservation of momentum, this is the same as the factor
obtained in equation (\ref{intermedone}).
In \cite{Mason:2008jy} a simplifying gauge choice was taken, where $\mu_{k}\propto\lambda_{n}$ for all $k=2,...,n-1$.
In this gauge, if we pass $P_{n-1}$ to the left so that it sits just
to the right of $\tilde{\mathcal{P}}_{1}$, the leading order contribution from the commutators (\ref{leadingpart})
vanishes because, via momentum conservation, this term is proportional to $v_{n-1}^{+}\ccdot p_{n}=0$.

So to evaluate the $P_{n-1}$ contribution, we first commute the $\omega_{\dot{b}}^{\dag}$ subfactor
to the left, where it hits the $\omega_{\dot a}$ oscillator in the $K_1$ term in ${\cal P}_1 = P_1 + K_1 \ell^{-2}$.
Using $[\omega_{\dot{b}}^{\dag},\omega_{\dot{a}}]=\varepsilon
_{\dot{a}\dot{b}}$, and that at large $N$, we can
replace $\pi^{\dag a}\pi_{b}=N\delta^{a}{}_{b}$, we obtain for the
subamplitude%
\begin{equation}
\frac{N}{\ellcc^{2}}\text{
}\left\langle x,\rho\right\vert V_{1}...V_{n}\tilde{\mathcal{P}}_{n}%
\hspace{1pt}\left\vert x,\rho\right\rangle \text{ }(v_{1}^{-}\!\cdot
\!\hspace{1pt}\hspace{1pt}v_{n-1}^{+})\text{ }\prod_{k=2}^{n-2}v_{k}^{+}%
\ccdot\left(  p_{1}+...+p_{k-1}\right)  .
\end{equation}
Notice the presence of the additional prefactor $N/\ellcc^{2}\simeq
1/\ell_{pl}^{2}$. The amplitude thus remains finite in the scaling limit
(\ref{scaling}). Including the overall normalization factor $\kappa^{n}$,
 the amplitude then scales as $\kappa^{n-2}$, as expected for the
MHV graviton amplitude.

The last step is to take into account the flow of momentum out of the patch,
as indicated by the factor of $\tilde{\mathcal{P}}_{n}$. The leading order
behavior is obtained by taking $\tilde{\mathcal{P}}_{n}\rightarrow P_{n}$,
which induces a small translation on the ket state $\left\vert x,\rho
\right\rangle \rightarrow\left\vert x+v_{n},\rho\right\rangle $. This has the
consequence that it shifts the domain of integration in equation
(\ref{projmink}). The effect of the shift is given by
the value of the subamplitude with $\left\vert x,\rho\right\rangle $ held fixed but with
$P_{n}$ acting to the left on $V_{1}$. This amounts to the substitution $V_{1}%
\rightarrow\left(  v_{n}\cdot p_{1}\right)  V_{1}$, and reflects the fact that
the minus helicity graviton backreacts on the patch as it flows into and then
out of the patch.

Now that we have taken into account the gravitational \textquotedblleft color
factor\textquotedblright, the remaining manipulations are the same as for
gluon scattering:%
\begin{align}
A_{1,...,n} & = \kappa^{n-2}\text{ } \text{ Tr}\left(  V_{1}\hspace
{1pt}...V_{n}\right)  \text{ }(v_{1}^{-}\ccdot\spc v_{n-1}^{+})\left(  v_{n}^{-}\ccdot\spc p_{1}\right)  \text{ }\prod
_{k=2}^{n-2}v_{k}^{+}\ccdot\left(  p_{1}+...+p_{k-1}\right)  \\
&  = \kappa^{n-2} \text{ }\text{ }\delta^{4|4}\times\frac{(v_{1}^{-}\!\cdot
\!\hspace{1pt}\hspace{1pt}v_{n-1}^{+})\left(  v_{n}^{-}\cdot p_{1}\right)
}{C(n)}\text{ }\prod_{k=2}^{n-2}v_{k}^{+}\ccdot\left(  p_{1}+...+p_{k-1}%
\right)
\end{align}
where the factor of $\delta^{4|4}$ enforces conservation of momentum on the
$\mathcal{N}=4$ superspace. Here, $C(n)=\langle12\rangle\langle23\rangle
\cdots\langle n\!-\!1\hspace{1pt}n\rangle\langle n1\rangle$ denotes the
Parke-Taylor denominator. In addition there are various multi-trace contributions from the fuzzy Hilbert space which
enter into both the Yang-Mills and gravity amplitudes. Following the arguments in section 7 of \cite{TMMII}, we expect these to
be related to Planck suppressed correction terms to an Einstein theory of gravity.

To complete the computation, we plug in the values of the reference
twistors used in our evaluation of the current algebra. For the plus helicity
polarization tensors, we have:%
\begin{equation}
\langle\hspace{1pt}\ast\hspace{1pt}\mu_{k}\rangle=\frac{\langle\hspace
{1pt}\ast\;n\rangle}{\langle kn\rangle}\label{dualsp}%
\end{equation}
for $k=2,...,n-1$. The dual spinor (\ref{dualsp}) automatically satisfies
$\langle\mu_{k}\lambda_{k}\rangle=1$. Note that this choice implies that
$\langle\mu_{k}\hspace{1pt}n\rangle=0$. The minus helicity polarization
tensors are:
\begin{equation}
\lbrack\hspace{1pt}\tilde{\mu}_{1}\hspace{1pt}\ast\hspace{1pt}]=\frac
{[\hspace{1pt}\beta\,\ast\hspace{1pt}]}{[\hspace{1pt}\beta\,1]},\qquad
\qquad\quad\lbrack\hspace{1pt}\tilde{\mu}_{1}\hspace{1pt}\ast\hspace
{1pt}]=\frac{[\hspace{1pt}\beta\,\ast\hspace{1pt}]}{[\hspace{1pt}\beta\,n]}%
\end{equation}
Inserting this explicit form of the polarization tensors, we obtain:%
\begin{equation}
A_{1,...,n} = \kappa^{n-2} \delta^{4}\left(
{\displaystyle\sum}
p_{i}\right) \text{ }\frac{ \langle 1n\rangle^{8}\,  \lbrack\hspace
{1pt}\beta\,n-1\hspace{1pt}]}{[\hspace{1pt}\beta\,n]\langle n-1n\rangle
\left\langle 1n\right\rangle ^{2}}\text{ } \frac{1}{C(n)} \text{ }\underset{k=2}{\overset{n-2}{%
{\displaystyle\prod}
}}\frac{[\hspace{1pt}k\hspace{1pt}|p_{1}+...+p_{k-1}|n\rangle}{\langle
k\hspace{1pt}n\rangle}%
\end{equation}
with $C(n) = \la 1 2 \ra \la2 3\ra \cdots \la n\!-\!1\spc  n \ra\la n 1\ra$
the usual Parke-Taylor denominator. Rather remarkably, this
is \textit{identical} to the form of the subamplitude
found in \cite{Mason:2008jy}! At this point, the combinatorics and
manipulations to show that this expression reproduces the original BGK result
(\ref{BGK}) are the same as in \cite{Mason:2008jy}. The full amplitude is now
obtained by summing over all permutations of the plus helicity gravitons
$2,...,n-1$. As explained in \cite{Mason:2008jy}, note that although the
amplitude appears to depend on the gauge choice $\beta$, once we sum over all
possible permutations, this dependence drops out, as it must. For formulae with a manifest $S_{n-2}$
permutation symmetry of all plus helicity gravitons see \cite{Nguyen:2009jk}.

\section{Conclusions} \label{sec:CONC}

In this paper we have studied the properties of a gaussian matrix model formulated on fuzzy twistor space.
The model possesses a number of symmetries which at large $N$ have a direct interpretation on a 4D space-time.
Identifying a natural class of currents for the theory, we have shown that correlators of the matrix model
reproduce both MHV gluon \emph{and} graviton scattering amplitudes. This provides non-trivial evidence that
the twistor matrix model of \cite{TMM, TMMII} correctly describes 4D physics, and moreover, contains a
gravitational subsector \cite{TMMII}.

Our discussion has been limited to MHV amplitudes. To recover more intricate amplitudes, we expect
that the details of bulk physics in twistor space will be important. This is in keeping with the qualitative
picture in twistor string theory that MHV gauge theory amplitudes localize on lines of twistor space \cite{Witten:2003nn}.
Let us note that the gravity sector adds an interesting twist to this; These amplitudes essentially localize
on a complex line, but one which has been deformed by the background gravitons. To make
further contact with 4D physics, it would be interesting to study more general types of
amplitudes in the framework proposed in \cite{TMMII}.

Since we have an explicit finite $N$ matrix model, we can also study various
$1/N$ corrections to the continuum theory result. In the flat space limit, the
corrections of physical relevance are controlled by the small parameter
$\ell_{pl}$. More broadly, one can also consider correlators on a finite
size $S^{4}$ which may provide a concrete means to
compute (via analytic continuation) in-in correlators on de Sitter space.

\section*{Acknowledgements}

We thank N. Arkani-Hamed, N. Berkovits, S. Caron-Huot, T. Hartman, C. Hull, D. Karabali, M.
Kiermaier, J. Maldacena, V.P. Nair, D. Skinner, C. Vafa, E. Verlinde and E. Witten for
helpful discussions. The work of JJH is supported by NSF grant PHY-0969448 and
by the William Loughlin membership at the Institute for Advanced Study. The work
of HV is supported by NSF grant PHY-0756966.


\def\txX{\tilde{\xX}}
\appendix
\section{Twistors and $SO(5)$}

In this Appendix we review some aspects of the infinity bitwistor for an $S^{4}$. See \cite{TMMII} for
a somewhat expanded discussion. With the help of the infinity twistor, we can define normalized six component
space-time coordinates $
\xX^{\alpha\beta}$, and their duals $\txX_{\alpha\beta}$, via
\be
\xX^{\alpha\beta} = \frac{Z^{[\alpha}W^{\beta]}} {\la\spc ZW\ra}, \qquad \quad
\txX_{\alpha\beta} = \frac 1 2 \epsilon_{\alpha\beta\gamma\delta} \xX^{\gamma
\delta}
\ee
These satisfy  the relations $\xX^{\alpha\gamma} \spc \xX_\gamma{}^{\beta} = \xX^{\alpha \beta},$
$\txX_{\alpha\gamma} \spc \txX^\gamma{}_{\beta} = \txX_{\alpha \beta},$.
The dual coordinates $\txX_{\alpha\beta}$ act as projection
matrices onto the twistor line associated to the space-time point $\xx$: the twistor
line equation 
\be
\label{twistorline}
\txX{\!\spc}_{\alpha\beta} {U}^\beta = 0
\ee
is solved by all points $U = a\spc Z + b\spc W$ on the twistor line through $Z$
and $W$.

The dual space-time coordinates $\txX{\!\spc}^{\spc \alpha}_{\; \beta}$ can be
parametrized as with the
help of five
coordinates $\xxn^{A}$  constrained to live on an $S^4$ of unit radius, via
\be
\label{xparam}
\qquad	\txX{\!\spc}_{\alpha\beta} =  \begin{pmatrix}
					\spc \frac 1 2 (1+  \xxn_5 ) \epsilon^{ab} \,
						& -    i  \xxn^a{}_{\dot b}\, \\[2mm]
 						\, {i}  \xxn_{\dot a}{}^{b} \, & \frac 1 2 (1-  {\xxn_5}  )
\epsilon_{\dot a \dot b}
					\end{pmatrix} , \qquad \quad \xxn^A \xxn_A = 1\, .
%
\ee
Here  $\xxn^{\; \dot a}_{b} =  \frac 1 2 (\xxn_\mu \sigma^\mu){}^{\; \dot a}_{b}$ with
$\sigma^{\mu} =
(\sigma^i,-i \mathbf{1})$  the usual Pauli matrices.
With this parametrization, the four component twistor line equation (\ref
{twistorline}) reduces to the
standard two-component
twistor line equation with
\be
{x^{\; \dot a}_{b}}= \, \frac {\spc 2 \xxn^{\; \dot a}_{b}}
{1 -  \xxn^5}\, .
\ee
The flat space limit amounts to zooming in on the south pole region of the $S^4$
near $\xxn^{\spc 5} \simeq -1$.
In this limit, the remaining four $S^4$ coordinates $\xxn^\mu$ become identified
with flat space coordinates $x^\mu$.

\section{Fuzzy Cauchy}

In this Appendix, we give an alternative construction of the dual position state $(\lambda|$
in ${\cal H}^*_{\mathbb{CP}^1}(N)$ that satisfies the inhomogeneous coherent state property (\ref{result}).

The dual Hilbert space  ${\cal H}^*_{\mathbb{CP}^1}(N)$ is $N+1$ dimensional. So let us introduce a convenient
position eigenbasis, by  picking a preferred set of $N+1$ positions $z_p$, with $p=0,.., N$, for which we
will then construct dual position eigenstates. We will then later define general position eigenstates for other
locations
via a suitable version of analytic continuation.
A natural choice is to take the special positions $z_p$ to lie on the unit circle
\be
z_p^{N+1} = 1, \qquad \text{so that} \qquad z_p = \exp\Bigl({\frac{ 2\ppi i p}{N+1}}
\Bigr).
\ee
We now introduce the discrete set of candidate dual position eigenstates, via
\be
\label{zstates}
\qquad \quad
(z_p | =\, \sum_{n=0}^{N} \, (\spc  n\spc | z_p^{n}  .
\ee
These form a complete basis of ${\cal H}^*_{\mathbb{CP}^1}(N)$. Moreover, they
are indeed almost dual position eigenstates. The violation of the coherent state
condition is localized at
the two poles:
\bea
\label{dualco}
(z_p | \bigl(\ppi_1 - z_p \ppi_2\bigr)  = (0| - (\infty|
\eea
where $(0|$ and $(\infty| = (N|$ denote the dual north and south pole state in $
{\cal H}^*_{\mathbb{CP}
^1}(N+1)$.

To extend the definition of the dual position states to other points away from the
unit circle, we will use a
discrete version
of Cauchy's formula. We introduce the following notation
\be
\oint_{\! z_p} \, (...)=  \frac{1}{N+1} \sum_{p=0}^{N} z_p\, (...)\quad
\ee
Indeed, we claim that the continuum limit of the right-hand side amounts to
performing a standard complex
contour integration along the
unit circle. As a first trivial check, we compute the basic Cauchy integral
\be
\label{cau}
\oint_{\! z_p}\, \frac {1} {z_p - \xi } = \frac{1}{1- \xi^{N+1}}   \, \equiv \, \theta_
\sN(\xi)
\ee
In the large $N$ limit, the function on the right hand side reduces to a step
function: it is equal to $1$ on
the northern hemisphere where $|\xi|< 1$,
and vanishes on the southern hemisphere where $|\xi|>1$.  So at large $N$
we can replace $\theta_\sN
(\xi)$ by
\be
\label{resultt}
\theta(\xi) =  \left\{ \begin{array}{cc} \, 1 \qquad  \text{for}\  |\xi| < 1 \\[3mm] 0
\qquad \text{for} \ |\xi| > 1
\end{array} \right.
\ee
This is indeed what one expects from a discretized residue theorem: when $\xi
$ traverses the
equator, it sneaks between the holes in the discrete contour and escapes.
\footnote{On the southern patch,
when $|\xi|<1$, we need to use the dual  coordinates $\lambda = \xi^{-1}$
and $w_p = z_p^{-1}$.
The Cauchy formula then becomes $\oint_{w_p} \frac{1}{w_p-\lambda} = \theta_
{\sN} (\lambda)$
where $\oint_{w_p} \! \!\!  = \oint_{z_p}\! \!  \frac{dw_p}{dz_p}$. }
We will return to this special feature of the discretized contour integration
momentarily.

With this new tool in hand, we now define dual position states for any position $
\xi$ via
\bea
\label{mudual}
(\xi | =  
\oint_{\! z_p}  \frac 1 {z_p - \xi} \; (z_p |  
\eea
This provides a well defined state for any value of $\xi$, except for solutions to
$\xi^{N+1} =~1$.
The state $(\xi|$ almost satisfies the dual coherent state requirement. Upon
inserting the definition (\ref
{mudual}), while using
equations (\ref{dualco}), (\ref{cau}) and the formula  {\large $\oint_{z_p}$}$(z_p|\pi_2 =
(\infty |,$
one finds after a straightforward calculation
\bea
\label{resultt}
(\xi| (\pi_1 - \xi \pi_2) =  \theta_\sN\! (\xi)\spc  (0 | \spc + \theta_\sN\! (\xi^
{-1})( \infty|
\eea
with $(0|$ and $(\infty|$ the dual south and north  pole state in ${\cal H}^*_
{\mathbb{CP}^1}(N+1)$.  In the
large $N$ limit, equation (\ref{result}) reduces to
\be
\label{resultt}
(\xi | (\pi_1 - \xi \pi_2) =  \left\{ \begin{array}{cc} \, (0| \qquad \  \text{for} \ |\xi|
< 1 \\[3mm](\infty| \qquad
\text{for} \ |\xi| > 1 \end{array} \right.
\ee
This is our desired intermediate result. It shows that the dual coherent state $
(\xi|$ are position eigenstates, modulo a source term localized at
the corresponding pole.

Finally, let us compute the overlap between the position eigenstates. A simple
calculation shows that the
dual basis $(z_p|$ pairs with the position states (\ref{lambdef}) via
\be
\label{overlap}
(z_p | \lambda) = 
\frac{1}{1 - z_p \spc \lambda}
\ee
Combining
this result with (\ref{mudual}),
we obtain that
\be
\label{mulamb}
( \spc \xi \spc | \spc \lambda \spc ) =
\oint_{\! z_p}   \frac 1 {(z_p - \xi)(1 - z_p \lambda)}  = \frac{ \theta(\xi)  - \theta
(\lambda^{-1})}{1- \xi
\lambda}
\ee
This equation reveals, as expected, that $\xi$ and $\lambda$ are reciprocal
affine coordinates. Note,
however, that the pole in the denominator is spurious.
The step functions do not allow $\xi$ and $\lambda$ to be located on the same
hemisphere: whenever
they do, the numerator vanishes.

\begin{figure}
[ptb]
\begin{center}
\includegraphics[
height=1.7253in,
width=1.7149in
]%
{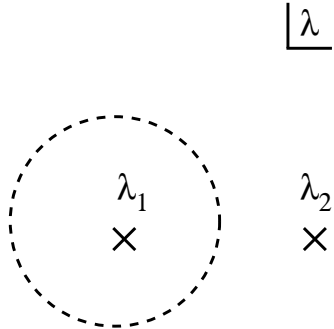}%
\caption{Evaluation of correlators of the chiral boson on a fuzzy
$\mathbb{CP}^{1}$ is achieved by working in terms of a basis of states which
vary holomorphically as a function of the complex plane parameter $\lambda$.
In the figure, two patches of the $\mathbb{CP}^{1}$ are indicated by the
dashed circular contour. Correlators in different patches of the
$\mathbb{CP}^{1}$ are computed by first evaluating in one patch in the large $N$ limit, and a
subsequent analytic continuation.}%
\label{contour}%
\end{center}
\end{figure}

The standard way to overcome this obstacle is via analytic continuation.
Consider the overlap $(\xi|
\lambda)$ and let $\lambda_1$ be the reciprocal coordinate  to $\xi_1 =
\lambda_1^{-1}$. We wish to {\it
define} the south patch state $(\lambda_1|$ via analytic continuation of the north
patch
state $(\xi_1|$. However, here we meet a subtlety. At infinite $N$, the step
functions are non-analytic at
the equator, reflecting the jump over the Cauchy contour. At {finite} $N$, on the
other hand, the step
functions $\theta_N(\xi)$  are perfectly analytic, reflecting that the contour is
just a discrete set of points.
The normal strategy of analytic continuation would amount to pushing the
location of the contour. Hence we can proceed in two ultimately equivalent ways.
We can either first take
the large $N$ limit, and then analytically continue. Or we can
define analytic continuation by suitably deforming the discrete contour. In
practice this means that
the location $\lambda_1 = \lambda_2$, where we expect the pole to occur,  must
lie somewhere on the
discrete contour. At very large $N$, this approaches the continuum prescription,
except very close to the
pole. At infinite $N$, the two prescriptions coincide.

The upshot is this: we {\it define} the state $(\lambda_1|$ such that its overlap
with $|\lambda_2)$ is given
by the analytic continuation of (\ref{mulamb}),
starting from the region $\theta(\xi) - \theta(\lambda^{-1}) = 1$. Hence in the
strict large $N$ limit, we have
\be
(\lambda_1 | \lambda_2) = \frac{1}{\lambda_1 - \lambda_2} \equiv \frac 1 {\la
\lambda_1 \lambda_2\ra}.
\ee
This completes our construction of the state $(\lambda|$.

\bibliographystyle{titleutphys}
\bibliography{GMM}

\end{document}